\documentclass[aps,prd,reprint,longbibliography,nofootinbib,floatfix,
author-year]{revtex4-2}

\usepackage[T1]{fontenc}
\usepackage{mathptmx}
\AtBeginDocument{\DeclareRobustCommand{\hbar}{{\mkern0.75mu\mathchar'26\mkern-9.75mu h}}}

\usepackage{graphicx}
\usepackage{float}

\usepackage{mathtools}
\usepackage{amssymb}
\usepackage{bm}
\usepackage{booktabs}
\usepackage{natbib}

\usepackage{xcolor}
\definecolor{darkblue}{RGB}{0,0,139}
\usepackage[colorlinks=true,linkcolor=darkblue,citecolor=magenta,urlcolor=darkblue]{hyperref}

\def\doi#1{\href{https://doi.org/#1}{\color{darkblue}#1}}

\newcommand{\rev}[1]{#1}
\newenvironment{revisedtext}{}{}

\usepackage{xstring}
\usepackage{tikz}
\usetikzlibrary{shapes.geometric,shapes.symbols}
\def\urlprefix{}
\makeatletter
\let\ORIGurl\url
\renewcommand{\url}[1]{%
	\begingroup
	\def\UrlArg{#1}%
	\IfBeginWith{\UrlArg}{https://doi.org/}{\href{\UrlArg}{\StrBehind{\UrlArg}{https://doi.org/}}}{%
		\IfBeginWith{\UrlArg}{http://doi.org/}{\href{\UrlArg}{\StrBehind{\UrlArg}{http://doi.org/}}}{%
			\IfBeginWith{\UrlArg}{https://arxiv.org/abs/}{\href{\UrlArg}{arXiv:\StrBehind{\UrlArg}{https://arxiv.org/abs/}}}{%
				\IfBeginWith{\UrlArg}{http://arxiv.org/abs/}{\href{\UrlArg}{arXiv:\StrBehind{\UrlArg}{http://arxiv.org/abs/}}}{%
					\IfBeginWith{\UrlArg}{https://ui.adsabs.harvard.edu/abs/}{\href{\UrlArg}{\StrBehind{\UrlArg}{https://ui.adsabs.harvard.edu/abs/}}}{%
						\IfBeginWith{\UrlArg}{http://ui.adsabs.harvard.edu/abs/}{\href{\UrlArg}{\StrBehind{\UrlArg}{http://ui.adsabs.harvard.edu/abs/}}}{%
							\ORIGurl{\UrlArg}%
	}}}}}}%
	\endgroup
}
\makeatother

\def\apj{ApJ}
\def\apjl{ApJ}

\def\aap{A\&A}

\def\prd{Phys.~Rev.~D}

\def\prl{Phys.~Rev.~Lett.}
\def\physrep{Phys.~Rep.}

\def\lrr{Living Rev.~Relativ.}

\def\zphy{Z.~Phys.}

\def\jhep{JHEP}

\def\pop{Phys.~Plasmas}
\def\cqg{Class.~Quantum Grav.}
\def\ijmpe{Int.~J.~Mod.~Phys.~E}

\newcommand{\mpsi}{m_\psi}
\newcommand{\mchi}{m_\chi}
\newcommand{\msig}{m_\sigma}
\newcommand{\momega}{m_\omega}
\newcommand{\gsig}{g_\sigma}
\newcommand{\gome}{g_\omega}
\newcommand{\Csig}{C_\sigma^2}
\newcommand{\Come}{C_\omega^2}
\newcommand{\Comemax}{C_{\omega,\mathrm{max}}^2}
\newcommand{\nsat}{n_{\mathrm{sat}}}
\newcommand{\Mpl}{M_{\mathrm{Pl}}}
\newcommand{\lamB}{\lambda_{\mathrm{B}}}

\newcommand{\Msun}{M_\odot}

\DeclareRobustCommand{\orcidicon}{%
	\begin{tikzpicture}
		\definecolor{orcidgreen}{HTML}{166B3A}
		\draw[orcidgreen, fill=orcidgreen] (0,0)
		circle [radius=0.16]
		node[white] {{\fontfamily{qag}\selectfont \tiny ID}};
	\end{tikzpicture}
	\hspace{-2mm}
}
\foreach \x in {A, ..., Z}{%
	\expandafter\xdef\csname orcid\x\endcsname{\noexpand\href{https://orcid.org/\csname orcidauthor\x\endcsname}{\noexpand\orcidicon}}
}

\begin{document}
	\raggedbottom

	\title{Self-bound quantum dark fermion stars: \\ saturation, a coupling-capped maximum mass,\\ and tidal signatures of a vector dark sector}

	\author{Il\'{i}dio Lopes\orcidA{}}
	\affiliation{\href{https://ror.org/020hp3377}{Centro de Astrof\'isica e Gravita\c{c}\~ao--CENTRA}, Departamento de F\'isica, \\ Instituto Superior T\'ecnico--IST, \href{https://ror.org/01c27hj86}{Universidade de Lisboa--UL}, \\ Av.~Rovisco Pais~1, 1049-001 Lisboa, Portugal}

	\date{June 2026}

\begin{abstract}
I show that a dark sector whose fermions couple to both a scalar and a vector mediator saturates in close analogy with nuclear matter: the energy per particle develops a minimum at finite density, cold dark matter of this kind is self-bound, and its stars, computed here in full general relativity, possess the sharp surfaces of strange quark stars.
Saturation makes the maximum stellar mass a property of the Lagrangian rather than an inheritance of the Chandrasekhar scaling: it rises eightfold across the saturation window, which a critical vector coupling closes, so the theory caps its own maximum mass, at $3.9\,M_\odot$ for a fermion of $1\,\mathrm{GeV}$\rev{ and the fiducial saturation density, a ceiling shared by the whole family of couplings rather than by any single parameter set}.
Metastable couplings support two disconnected stable families, one dilute and gravitationally bound, the other dense and self-bound.
The quantum-pressure correction, confined by saturation to a surface layer a few Compton wavelengths thick, governs the smallest dark nuggets whilst shifting stellar masses by less than one part in $10^{21}$.
A $1.4\,M_\odot$ member of the fiducial family has a radius of $13.3\,\mathrm{km}$\rev{, a compactness of $0.155$,} and a tidal deformability of $1.6\times 10^{3}$, almost an order of magnitude above the measured neutron-star value, so \rev{at a given particle mass and couplings} a single well-measured inspiral
 distinguishes these stars from neutron stars and from their scalar-only counterparts, whilst the deformability surviving at the maximum mass separates them, in principle, from black holes.
\end{abstract}

	\maketitle

	\section{Introduction}
	\label{sec:intro}

	The proposition that part of the dark matter resides in compact, self-gravitating objects has matured from speculation into a programme with sharply defined observational targets \citep{2024PhR..1052....1B,2019LRR....22....4C,2018PhR...730....1T}.
	If the dark sector contains stable fermions endowed with their own mediator fields, then the same physics that assembles nucleons into nuclei and neutron stars operates, with rescaled couplings and masses, on the dark side of the Lagrangian: short-range attraction binds the matter, degeneracy pressure and short-range repulsion support it, and gravity supplies the long-range confinement that turns bound matter into stars \citep{2014PhRvD..90e5030W,2017PhRvD..96i6012G,2006PhRvD..74f3003N,2015PhRvD..92f3526K}.
	The observational stakes are considerable: such objects populate the same mass and radius windows as neutron stars and low-mass black holes, and gravitational-wave interferometers now measure the very quantities that distinguish one population from another, namely masses, radii, and tidal deformabilities \citep{2017PhRvL.119p1101A,2018PhRvL.121p1101A,2021ApJ...918L..28M}.

	This paper extends a systematic programme on quantum dark fermion stars, configurations in which the Bohm quantum-pressure correction enters the equilibrium equations on the same footing as gravity and degeneracy pressure.
	The Newtonian foundations were laid in \citet{LopesQuantumStarsNewtonian2026}; the present work shares with that study the Bohm closure and the two-fluid formalism, promoted here to general relativity, where the structure problem becomes the two-fluid Tolman--Oppenheimer--Volkoff (TOV) system, and to an entirely distinct matter sector.
	The scalar-only theory produces what one might call soft stars: gravitationally bound, comparatively dilute configurations whose edges are set by the quantum-pressure tail and whose maximum masses descend, in essence, from the Chandrasekhar scaling.
	The question addressed here is what becomes of these objects when the dark sector carries, in addition, a neutral vector field coupled to the conserved fermion number current, the dark analogue of the $\omega$ meson of relativistic nuclear mean-field theory \citep{1974AnPhy..83..491W,1997IJMPE...6..515S,1977NuPhA.292..413B}.

	The answer, as we shall see, is fourfold, and none of it is accessible to the scalar-only theory: competition between the two mediator ranges generates dark-sector saturation, so the matter is self-bound, a degenerate analogue of the dark nuclei of \citet{2014PhRvD..90e5030W} and \citet{2017PhRvD..96i6012G}, with the Bohm term compressed from a broad tail into a microscopically thin surface layer; the maximum mass becomes a falsifiable function of the couplings $(\gsig,\gome,\msig,\momega)$; the equation of state stiffens, raising radii and tidal deformabilities at fixed mass; and metastable couplings open a second stable branch, the dark analogue of the white-dwarf and neutron-star families.
	Relativistic fermion stars with an effective repulsive interaction were charted by \citet{2006PhRvD..74f3003N}, and asymmetric dark stars bound by Yukawa self-interactions by \citet{2015PhRvD..92f3526K}; the present construction departs from both already for a single species, in that scalar--vector competition renders the matter self-bound rather than merely stiffened, the saturation window caps the maximum mass through the couplings, and a quantum surface term governs the smallest configurations, whilst the two-species equilibrium closure generalises the formalism without underpinning any of these results.

	The principal original contributions of this work are:
	(i)~the curved-spacetime $\sigma$--$\omega$ field equations for two dark fermion species and their reduction to a thermodynamically consistent relativistic mean-field equation of state;
	(ii)~the species-wise force balance, derived in full from the field equations, and the equilibrium closure $\mu_i\,e^{\nu}=\mathrm{const}$ extended to include both the vector potential and the Bohm correction evaluated with the effective mass;
	(iii)~the dark saturation curve and the self-bound stellar surface, together with the maximal vector coupling, $\Come\simeq 512$ at the fiducial saturation density, beyond which saturation cannot occur;
	(iv)~the maximum-mass map $M_{\rm max}(\Csig,\Come)$, ascending across the saturation window from below the Oppenheimer--Volkoff value \citep{1939PhRv...55..374O,2006PhRvD..74f3003N} to $2.41\,\Mpl^3/m_1^2$;
	(v)~the two disconnected stable branches of metastable matter;
	(vi)~the Bohm surface analysis, comprising the surface tension, the minimum nugget number, and the (reassuringly negligible) stellar-mass correction; and
	(vii)~the tidal-deformability discriminants separating the $\sigma$--$\omega$ star from the scalar-only star, from neutron stars, and from black holes.

	The remainder of the paper is organised as follows.
	Section~\ref{sec:model} presents the Lagrangian and the curved-spacetime field equations.
	Section~\ref{sec:eos} constructs the mean-field thermodynamics, the two-fluid TOV--Bohm system, and the modified closure, and fixes the dimensionless formulation.
	Section~\ref{sec:saturation} explores dark-sector saturation and the equation of state; Section~\ref{sec:stars} presents the stellar structure, the maximum-mass map, and the branch structure; Section~\ref{sec:surface} analyses the Bohm surface layer and the two-species configurations; and Section~\ref{sec:observations} translates the results into gravitational-wave observables.
	Section~\ref{sec:conclusions} concludes.
	Throughout, I adopt natural units ($\hbar=c=1$) with the metric signature $(-,+,+,+)$ \citep{1973grav.book.....M}, restoring physical units where clarity demands; the non-reduced Planck mass is $\Mpl=G^{-1/2}$.

	\section{Lagrangian and field equations}
	\label{sec:model}

	In this section I state the matter content and derive the field equations whose static solutions govern the stellar interior.
	The system comprises two dark spin-$\tfrac{1}{2}$ species, $\psi$ (mass $m_1\equiv\mpsi$) and $\chi$ (mass $m_2\equiv\mchi$), a real scalar mediator $\sigma$ (mass $\msig$), and a neutral vector mediator $\omega_\mu$ (mass $\momega$); the fermions carry no direct contact interactions, all binding being orchestrated by the two mediators and by gravity.

	\subsection{The dark-sector Lagrangian in curved spacetime}
	\label{sec:lagrangian}

	The action is $S=\int\mathrm{d}^4x\,\sqrt{-g}\,\mathcal{L}$, where the matter Lagrangian is the curved-spacetime counterpart of the $\sigma$--$\omega$ Lagrangian of quantum hadrodynamics \citep{1974AnPhy..83..491W,1997IJMPE...6..515S},
	\begin{equation}
		\begin{aligned}
			\mathcal{L} ={}& \sum_{i=1,2}\bar\psi_i\bigl(i\gamma^\mu D_\mu - m_i + g_{\sigma i}\,\sigma - g_{\omega i}\,\gamma^\mu\omega_\mu\bigr)\psi_i \\
			&-\frac{1}{2}\,g^{\mu\nu}\partial_\mu\sigma\,\partial_\nu\sigma - U(\sigma) \\
			&-\frac{1}{4}\,\Omega_{\mu\nu}\Omega^{\mu\nu} - \frac{1}{2}\momega^2\,\omega_\mu\omega^\mu \,,
		\end{aligned}
		\label{eq:Lagrangian}
	\end{equation}
	where $\psi_1\equiv\psi$, $\psi_2\equiv\chi$, $D_\mu=\partial_\mu+\Gamma_\mu$ is the spinor covariant derivative \citep{1996qtf..book.....B,2009qftc.book.....P}, $\Omega_{\mu\nu}=\nabla_\mu\omega_\nu-\nabla_\nu\omega_\mu$ is the vector field strength, and the scalar potential carries the most general renormalisable self-interactions in the Boguta--Bodmer form \citep{1977NuPhA.292..413B},
	\begin{equation}
		U(\sigma)=\frac{1}{2}\msig^2\sigma^2
		+\frac{b}{3}\,m_1\bigl(\gsig\sigma\bigr)^3
		+\frac{c}{4}\bigl(\gsig\sigma\bigr)^4\,,
		\label{eq:Usigma}
	\end{equation}
	with $\gsig\equiv g_{\sigma 1}$; the dimensionless self-couplings $b$ and $c$ carry the traditional symbols of the nuclear problem, the latter not to be confused with the speed of light, which natural units set to unity, and $c>0$ ensures boundedness from below.
	The vector mass is taken as given; it may be regarded as the relic of a spontaneously broken dark $U(1)$ whose Higgs mode is too heavy to figure in what follows, the Lagrangian above being the effective theory below that scale.
	Gravity is described by the Einstein--Hilbert term, and the static, spherically symmetric line element takes the familiar form
	\begin{equation}
		\mathrm{d}s^2 = -e^{2\nu(r)}\,\mathrm{d}t^2 + \left(1-\frac{2Gm(r)}{r}\right)^{\!-1}\mathrm{d}r^2 + r^2\,\mathrm{d}\Omega^2\,.
		\label{eq:metric}
	\end{equation}
	The scalar coupling enters the Dirac operator through the combination $m_i-g_{\sigma i}\sigma$, so each species propagates with the effective mass
	\begin{equation}
		m_i^*(r) = m_i - g_{\sigma i}\,\sigma(r)\,,
		\label{eq:mstar}
	\end{equation}
	and the scalar channel is attractive wherever $\sigma>0$; the vector coupling, by contrast, enters through the conserved current $j_i^\mu=\bar\psi_i\gamma^\mu\psi_i$ and is repulsive between like charges, as in the nuclear prototype \citep{1974AnPhy..83..491W,1997IJMPE...6..515S}.
	It is worth remarking that this sign structure is not a modelling choice but a structural feature of the exchange amplitude: scalar exchange between identical fermions is attractive, and vector exchange between like charges repulsive.

	Variation of the action yields the curved-spacetime field equations, the covariant lifts of the meson field equations of the nuclear problem \citep{1977NuPhA.292..413B,1997csnp.book.....G}.
	The scalar obeys the nonlinear Klein--Gordon equation
	\begin{equation}
		\begin{split}
			\Box_g\,\sigma - \msig^2&\sigma - b\,m_1\gsig\bigl(\gsig\sigma\bigr)^2 - c\,\gsig\bigl(\gsig\sigma\bigr)^3 \\
			&= -\sum_i g_{\sigma i}\,\langle\bar\psi_i\psi_i\rangle\,,
		\end{split}
		\label{eq:KG_sigma}
	\end{equation}
	whose source is the Lorentz-scalar density, and the vector obeys the Proca equation \citep{Proca1936},
	\begin{equation}
		\nabla_\mu\Omega^{\mu\nu} - \momega^2\,\omega^\nu = \sum_i g_{\omega i}\,\langle\bar\psi_i\gamma^\nu\psi_i\rangle\,.
		\label{eq:Proca}
	\end{equation}
	For the static configurations of interest only the time component $\omega_0$ survives, the spatial components vanishing with the three-current, and Eq.~\eqref{eq:Proca} reduces to a screened equation that slots in alongside the scalar one,
	\begin{equation}
		\bigl(\nabla^2_{\mathrm{cov}} - \momega^2\bigr)\,\omega_0 = -\sum_i g_{\omega i}\,n_i\,,
		\label{eq:omega_screened}
	\end{equation}
	where $\nabla^2_{\mathrm{cov}}$ is the covariant Laplacian of the metric~\eqref{eq:metric} acting on the static potential $\omega_0$ and $n_i$ is the local number density of species~$i$; the curvature terms that distinguish the vector operator from its scalar counterpart are suppressed by the square of the ratio of the mediator Compton wavelength to the stellar radius and play no role in what follows.
	The conservation of each $j_i^\mu$ guarantees that the vector source is the number density itself, with no relativistic suppression; the scalar source, by contrast, is the scalar density $n_{s,i}$, which falls below $n_i$ as the matter becomes relativistic.
	This asymmetry between the two sources, attraction that weakens with density and repulsion that does not, is the microscopic origin of saturation, and the reader will recognise it as the very mechanism that prevents ordinary nuclei from collapsing.

	\subsection{Mean-field reduction}
	\label{sec:meanfield}

	Two small parameters justify the approximations adopted in the remainder of the paper.
	First, the mediator Compton wavelengths $1/\msig$ and $1/\momega$ are microscopic, of order a few times $1/m_1$, whereas the stellar radius is of order $\Mpl/m_1^2$, larger by the enormous factor $\Mpl/m_1$; the gradient terms in Eqs.~\eqref{eq:KG_sigma} and \eqref{eq:omega_screened} are therefore negligible everywhere except within a few Compton wavelengths of the surface, and the fields track their local sources algebraically.
	This is the Thomas--Fermi, or local-density, regime, and it is the same approximation that underlies relativistic mean-field nuclear structure \citep{1997IJMPE...6..515S,1997csnp.book.....G}.
	Secondly, in the mean-field approximation the bilinears are replaced by their ground-state expectation values, $\langle\bar\psi_i\psi_i\rangle\to n_{s,i}$ and $\langle\bar\psi_i\gamma^0\psi_i\rangle\to n_i$, an approximation whose nuclear pedigree extends from \citet{1974AnPhy..83..491W} to modern energy-density functionals.
	The no-sea prescription is adopted in the same breath, the Dirac-sea contribution to the sources being discarded as in the nuclear prototype; the neglect is least innocent where the effective mass approaches zero, a regime entered below only by the scalar-only control.
	The surface region, where the local-density approximation fails and the Bohm gradient correction dominates, receives a dedicated treatment of its own in due course.

	Within this regime the meson field equations become algebraic,
	\begin{align}
		\msig^2\sigma + b\,m_1\gsig^3\sigma^2 + c\,\gsig^4\sigma^3 &= \sum_i g_{\sigma i}\,n_{s,i}\,,
		\label{eq:sigma_alg}\\
		\momega^2\,\omega_0 &= \sum_i g_{\omega i}\,n_i\,,
		\label{eq:omega_alg}
	\end{align}
	and the entire equation of state depends on the couplings and mediator masses only through the two dimensionless combinations familiar from the Walecka model,
	\begin{equation}
		\Csig \equiv \gsig^2\,\frac{m_1^2}{\msig^2}\,,\qquad
		\Come \equiv \gome^2\,\frac{m_1^2}{\momega^2}\,,
		\label{eq:CsCv}
	\end{equation}
	with $\gome\equiv g_{\omega 1}$.
	\rev{The saturation properties therefore depend on the couplings and mediator masses only through these combinations, never on $\gsig$, $\gome$, $\msig$, or $\momega$ separately: a stronger coupling is exactly compensated by a heavier mediator. Nor do the two combinations act independently: the saturation density and the binding fraction together fix the pair $(\Csig,\Come)$ (Section~\ref{sec:satfit}), and a deeper binding at fixed saturation density requires both to grow in concert, the additional scalar attraction being held in check by additional vector repulsion (Table~\ref{tab:sets}).}
	Unless stated otherwise I take universal couplings, $g_{\sigma 2}=g_{\sigma 1}$ and $g_{\omega 2}=g_{\omega 1}$, a restriction that deserves a precise statement, the formalism above being fully general whilst the survey below is a deliberate slice.
	At its most general the theory carries seven dimensionless parameters, the two coupling pairs $g_{\sigma i}^2\,m_1^2/\msig^2$ and $g_{\omega i}^2\,m_1^2/\momega^2$, the mass ratio $m_2/m_1$, and the self-couplings $(b,c)$; moreover, because each number current is separately conserved, the composition $N_2/N_1$ is a further free function rather than an output of equilibrium.
	I restrict attention to universal couplings and a single dominant species in order to isolate the saturation physics that drives every result of this paper, the species-dependent machinery being exercised once below, for a heavy admixed minority.
	Throughout, the dark sector is taken to be secluded: neither mediator couples to Standard-Model fields, so direct-detection and collider searches place no constraint on the combinations~\eqref{eq:CsCv}, which answer only to the internal consistency requirements of the theory itself; the corresponding price is that gravity, and gravitational waves in particular, supply the only observational access to these objects.
	Seclusion exacts a further price: the particle mass $m_1$ cannot then be supplied by laboratory measurement, so wherever a result below is anchored to a known $m_1$, that knowledge must descend from the broader dark-matter context; questions of cosmological formation and abundance, which would fix it, lie outside the scope of this paper, whose concern is the structure and signatures of the objects once formed.

	\section{Equation of state, equilibrium structure, and closure}
	\label{sec:eos}

	\subsection{Mean-field thermodynamics}
	\label{sec:thermo}

	At zero temperature each species fills a Fermi sea of momentum $p_{F,i}=(3\pi^2 n_i)^{1/3}$ with single-particle energies $\sqrt{p^2+m_i^{*2}}$.
	The kinetic contributions to the energy density and pressure are the standard relativistic integrals
	\begin{align}
		\varepsilon_{\mathrm{kin},i} &= \frac{1}{\pi^2}\int_0^{p_{F,i}}\!\sqrt{p^2+m_i^{*2}}\;p^2\,\mathrm{d}p\,,
		\label{eq:eps_kin}\\
		P_{\mathrm{kin},i} &= \frac{1}{3\pi^2}\int_0^{p_{F,i}}\!\frac{p^4}{\sqrt{p^2+m_i^{*2}}}\,\mathrm{d}p\,,
		\label{eq:P_kin}
	\end{align}
	and the scalar density sourcing Eq.~\eqref{eq:sigma_alg} is
	\begin{equation}
		n_{s,i} = \frac{1}{\pi^2}\int_0^{p_{F,i}}\!\frac{m_i^*}{\sqrt{p^2+m_i^{*2}}}\;p^2\,\mathrm{d}p\,,
		\label{eq:ns}
	\end{equation}
	all three possessing closed forms that the numerical implementation employs; in the free-gas limit, $m_i^*\to m_i$ with vanishing couplings, they collapse to the classical parametric equation of state of the relativistic Fermi gas \citep{1983bhwd.book.....S}.
	Collecting kinetic and field contributions, the total energy density and pressure of uniform matter read
	\begin{align}
		\varepsilon &= \sum_i \varepsilon_{\mathrm{kin},i} + U(\sigma) + \frac{1}{2}\momega^2\,\omega_0^2\,,
		\label{eq:eps_tot}\\
		P &= \sum_i P_{\mathrm{kin},i} - U(\sigma) + \frac{1}{2}\momega^2\,\omega_0^2\,,
		\label{eq:P_tot}
	\end{align}
	with $\sigma$ and $\omega_0$ determined by Eqs.~\eqref{eq:sigma_alg} and \eqref{eq:omega_alg} at each density, and the chemical potentials are
	\begin{equation}
		\mu_i = \sqrt{p_{F,i}^2 + m_i^{*2}} + g_{\omega i}\,\omega_0\,.
		\label{eq:mu_i}
	\end{equation}
	The opposite signs with which $U(\sigma)$ enters Eqs.~\eqref{eq:eps_tot} and \eqref{eq:P_tot}, against the like signs of the vector term, encode the entire phenomenology: scalar condensation lowers the pressure whilst the vector condensate raises it.
	The construction satisfies the thermodynamic identity $P=\sum_i\mu_i n_i-\varepsilon$ exactly; the numerical implementation verifies it to machine precision ($\lesssim 10^{-15}$ in relative terms), which provides a stringent internal check on the field equations and their solution.

	\subsection{Two-fluid TOV--Bohm system and the modified closure}
	\label{sec:tov}

	The macroscopic structure follows from covariant conservation of each species' stress-energy tensor under the shared metric~\eqref{eq:metric}.
	The Einstein equations supply the mass and potential equations,
	\begin{equation}
		\frac{\mathrm{d}m}{\mathrm{d}r} = 4\pi r^2\varepsilon\,,\qquad
		\frac{\mathrm{d}\nu}{\mathrm{d}r} = \frac{G\bigl(m+4\pi r^3 P\bigr)}{r\bigl(r-2Gm\bigr)}\,,
		\label{eq:mass_nu}
	\end{equation}
	and each species obeys the quantum-corrected force balance derived in Appendix~\ref{app:tov},
	\begin{equation}
		\frac{\mathrm{d}P_i}{\mathrm{d}r} = -(\varepsilon_i+P_i)\,\frac{\mathrm{d}\nu}{\mathrm{d}r} - n_i\,\frac{\mathrm{d}Q_i^{\mathrm{GR}}}{\mathrm{d}r}\,,
		\label{eq:TOV_bohm}
	\end{equation}
	where the curved-spacetime Bohm potential is appended at the fluid level with the effective mass in the denominator,
	\begin{equation}
		Q_i^{\mathrm{GR}} = -\frac{\lamB}{2m_i^*}\,\frac{\nabla^2_{\mathrm{cov}}\sqrt{n_i}}{\sqrt{n_i}}\,,
		\label{eq:bohm}
	\end{equation}
	with $\lamB=1/9$ the Kirzhnits coefficient, arising from the gradient expansion of the kinetic functional \citep{1935ZPhy...96..431W,1957JETP....5...64K,2018PhPl...25c1903M}.
	The value of $\lamB$ deserves a word: the single-orbital coefficient of \citet{1935ZPhy...96..431W} is unity, the value appropriate to a coherent condensate, and antisymmetrisation of the filled Fermi sea dilutes the gradient response ninefold \citep{1957JETP....5...64K}, a statement equivalent to requiring that the functional reproduce the static, long-wavelength limit of the Lindhard response; the coefficient is regime-dependent, rising towards unity for dynamic perturbations \citep{2018PhPl...25c1903M}, but a hydrostatic star probes precisely the static limit in which $1/9$ is the exact answer.
	Since the thickness and tension of the surface layer scale as $\sqrt{\lamB}$, the ninefold dilution is a factor the observables feel rather than a refinement of principle.
	The replacement $m_i\to m_i^*(r)$ in Eq.~\eqref{eq:bohm} is the natural prescription in a medium where the carriers of the gradient energy are quasiparticles of effective mass $m_i^*$; since $m_i^*<m_i$ throughout the interior, the quantum pressure is everywhere somewhat enhanced relative to the bare-mass prescription.
	One caveat deserves record: the coefficient $\lamB=1/9$ descends from the gradient expansion of the non-relativistic kinetic functional, and its relativistic generalisation acquires corrections that grow with $p_F/m^*$; at the surface, where alone the Bohm term matters, the fiducial matter has $p_F/m^*\simeq 0.6$, so these corrections are modest, whilst in the deep interior the term is negligible whatever its coefficient.

	Integrating Eq.~\eqref{eq:TOV_bohm} radially, with the chemical potential~\eqref{eq:mu_i} supplying the integrating factor, yields the algebraic first integral that closes the system: for each species,
	\begin{equation}
		\Bigl[\sqrt{p_{F,i}^2+m_i^{*2}} + g_{\omega i}\,\omega_0 + Q_i^{\mathrm{GR}}\Bigr]e^{\nu(r)} = \mu_i^\infty\,,
		\label{eq:closure}
	\end{equation}
	a constant throughout the configuration; a self-contained derivation is given in Appendix~\ref{app:closure}.
	Equation~\eqref{eq:closure} is the Tolman--Klein condition \citep{1930PhRv...35..904T,1949RvMP...21..531K} generalised twice over: the vector potential $g_{\omega i}\,\omega_0$ joins the kinetic term inside the bracket, exactly as the electrostatic potential joins the chemical potential of charged matter, and the Bohm potential $Q_i^{\mathrm{GR}}$ joins it on the same footing.
	In the Newtonian limit ($e^\nu\to 1+\Phi_g$) the closure reduces to $\mu_i = h_i(n_i;m_i^*) + m_i\Phi_g + g_{\omega i}\,\omega_0 + Q_i$ with $h_i$ the kinetic enthalpy per particle, which, once the vector term is switched off, is the form employed in the Newtonian instalment of this programme \citep{LopesQuantumStarsNewtonian2026}.

	Away from the thin surface layer the Bohm term is negligible, and the system reduces to the standard relativistic structure problem.
	For single-species configurations I integrate the TOV equation with the tabulated barotrope $\varepsilon(P)$ of Section~\ref{sec:thermo}; for two-species configurations I integrate the metric functions $(m,\nu)$ and recover the local densities $(n_1,n_2)$ at each radius by inverting the pair of closure conditions~\eqref{eq:closure}, an inversion that automatically locates the distinct radii at which each species is exhausted.

	\subsection{Dimensionless formulation and physical scales}
	\label{sec:units}

	The microphysics depends only on $(\Csig,\Come,b,c)$ once densities are expressed in units of $m_1$, and gravity introduces the single scale $\Mpl/m_1^2$.
	I therefore quote energy densities and pressures in units of $m_1^4$, radii in units of
	\begin{equation}
		\mathcal{R}_\ast = \frac{\Mpl}{m_1^2} = 2.409\;\Bigl(\frac{m_1}{\mathrm{GeV}}\Bigr)^{\!-2}\,\mathrm{km}\,,
		\label{eq:length_unit}
	\end{equation}
	and masses in units of
	\begin{equation}
		\mathcal{M}_\ast = \frac{\Mpl^3}{m_1^2} = 1.631\;\Bigl(\frac{m_1}{\mathrm{GeV}}\Bigr)^{\!-2}\,M_\odot\,,
		\label{eq:mass_unit}
	\end{equation}
	so that every dimensionless result in Sections~\ref{sec:saturation}--\ref{sec:surface} converts to physical units through the single parameter $m_1$.
	The TOV system in these units carries no free constants, and the surface compactness $\kappa=GM/R_T$, with $R_T$ the truncation radius, in the nomenclature of this programme, the stellar radius at which the pressure vanishes, is invariant.
	As a benchmark, the free Fermi gas ($\Csig=\Come=0$) yields a maximum mass of $0.3842\,\Mpl^3/m_1^2$ in the present implementation, in agreement with the classic Oppenheimer--Volkoff value $0.384\,\Mpl^3/m_1^2$ \citep{1939PhRv...55..374O,2006PhRvD..74f3003N}; this, together with the thermodynamic identity quoted above and the Buchdahl bound \citep{1959PhRv..116.1027B} respected by every configuration reported below (a necessary, if undemanding, check), constitutes the validation suite of the solver.

	The boundary-value problem is solved by adaptive Runge--Kutta integration, implemented with the SciPy library \citep{2020NatMe..17..261V}, of Eq.~\eqref{eq:mass_nu} from a regular centre, $m(0)=0$ with the central pressure the single free datum labelling the sequence, to the surface $P=0$, which for self-bound matter occurs at finite density; the potential $\nu$, fixed by Eq.~\eqref{eq:mass_nu} only up to a constant, is normalised a posteriori by matching to the exterior Schwarzschild solution, $e^{2\nu(R_T)}=1-2GM/R_T$, which is also what anchors the redshifted constants of the closure~\eqref{eq:closure}; equilibrium sequences are generated by scanning the central density, and the tidal properties reported below follow from the standard quadrupolar perturbation equation integrated alongside the background \citep{2008ApJ...677.1216H,2008PhRvD..77b1502F}.

	\section{Dark-sector saturation and the equation of state}
	\label{sec:saturation}

	\subsection{Saturation and parameter sets}
	\label{sec:satfit}

	Before any star can be built, the bulk matter must be characterised, and the central question is whether the energy per particle $E/A=\varepsilon/n$ possesses a minimum at finite density.
	\rev{The density at which that minimum occurs is the saturation density, $\nsat$; since $P=n^{2}\,\mathrm{d}(E/A)/\mathrm{d}n$, it is equivalently the non-zero density at which the pressure of uniform matter vanishes with positive incompressibility, so that matter at $\nsat$ is in mechanical equilibrium with the vacuum, and $(E/A)_{\mathrm{sat}}$ is the energy per particle of the most bound uniform state.}
	I fix the fiducial saturation point by the dimensionless Fermi momentum $x_{\mathrm{sat}}\equiv p_F/m_1=0.30$, corresponding to the number density $\nsat=9.12\times 10^{-4}\,m_1^3$, which for $m_1=1\,\mathrm{GeV}$ evaluates to $0.119\,\mathrm{fm}^{-3}$, agreeably close to the nuclear value of $0.16\,\mathrm{fm}^{-3}$ and a deliberate echo of it.
	Given a target binding fraction $B\equiv 1-(E/A)_{\mathrm{sat}}/m_1$, the two conditions $P(\nsat)=0$ and $(E/A)_{\mathrm{sat}}=(1-B)\,m_1$ determine $(\Csig,\Come)$ uniquely at fixed $(b,c)$; since $P=n^{2}\,\mathrm{d}(E/A)/\mathrm{d}n$ \citep{1983bhwd.book.....S}, the first condition places the minimum of the energy per particle at $\nsat$ and the second fixes its depth, exactly as the empirical saturation point determines the couplings of the nuclear Walecka model \citep{1974AnPhy..83..491W,1997csnp.book.....G}.

	\begin{table}
	\centering
	\caption{Parameter sets used throughout this work. All sets saturate at $x_{\mathrm{sat}}=0.30$ ($\nsat=9.12\times10^{-4}\,m_1^3$) except set~E, which shares the scalar coupling of set~A but carries no vector channel and saturates instead at its own far deeper coexistence point, near $5.9\,\nsat$. $B$ is the binding fraction at saturation, $K=9\,\mathrm{d}P/\mathrm{d}n|_{\mathrm{sat}}$ the incompressibility, and $m^*/m$ the effective-mass ratio at saturation.}
	\label{tab:sets}
	\begin{tabular}{@{}lcccccc@{}}
		\toprule
		Set & $\Csig$ & $\Come$ & $b$ & $c$ & $B$ & $K\,[m_1]$;\ $m^*/m$ \\
		\midrule
		A (fiducial) & $553.2$ & $400.8$ & $0$ & $0$ & $0.02$ & $0.672$;\ $0.536$ \\
		B (deep)     & $734.3$ & $481.4$ & $0$ & $0$ & $0.05$ & $1.490$;\ $0.414$ \\
		C (shallow)  & $395.3$ & $296.4$ & $0$ & $0$ & $0.005$ & $0.258$;\ $0.660$ \\
		D (soft)     & $340.3$ & $154.0$ & $0.003$ & $0.010$ & $0.02$ & $0.334$;\ $0.784$ \\
		E (scalar only) & $553.2$ & $0$ & $0$ & $0$ & -- & -- \\
		\bottomrule
	\end{tabular}
	\end{table}

\begin{figure}
	\centering
	\includegraphics[width=\columnwidth]{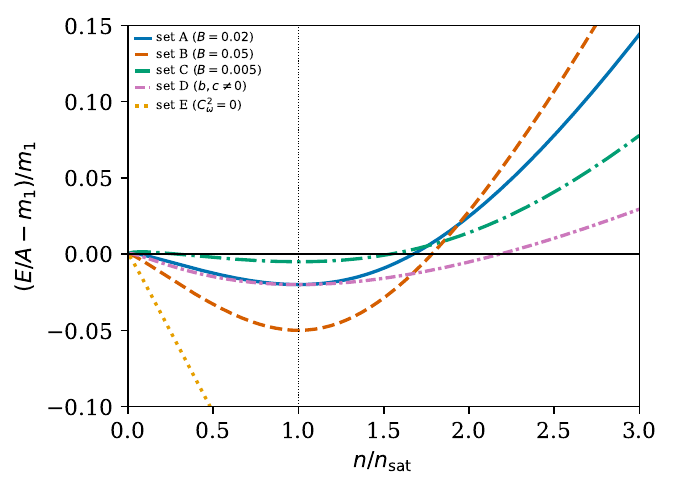}
	\caption{Energy per particle, $(E/A-m_1)/m_1$, as a function of density for the parameter sets of Table~\ref{tab:sets}. Sets A--D saturate at $n=\nsat$ with binding fractions $B=0.02$, $0.05$, $0.005$, and $0.02$ respectively; the scalar-only set~E (dotted) possesses no minimum in the displayed window, descending towards a far deeper minimum near $5.9\,\nsat$. The vertical dotted line marks $n=\nsat$.}
	\label{fig:saturation}
	\end{figure}

	Table~\ref{tab:sets} collects the five parameter sets explored below: a fiducial binding of two per cent (set~A), a deep and a shallow variant (sets~B and~C, the former deliberately stiff so as to bracket the family from above), a set in which the cubic and quartic self-couplings soften the incompressibility at fixed binding (set~D), and the scalar-only control (set~E) obtained from set~A by switching off the vector channel.
	The corresponding saturation curves appear in Fig.~\ref{fig:saturation}.
	Sets~A--D display the canonical structure, a minimum of depth $B\,m_1$ at $\nsat$, whereas the scalar-only curve plunges monotonically to far deeper binding at several times $\nsat$: with nothing to oppose the scalar attraction except the relativistic quenching of its own source, the matter saturates only when $m^*$ has collapsed nearly to zero, at a density $n\simeq 5.9\,\nsat$ and a binding more than an order of magnitude deeper.
	The vector channel, in other words, does not merely adjust the saturation point; it relocates the entire equilibrium of the theory.

	Two remarks on the couplings are in order.
	First, the fitted values $\Csig\sim 340$--$730$ and $\Come\sim 150$--$480$ sit in, and for the deeper bindings somewhat above, the range occupied by their nuclear counterparts (\rev{$\Csig=357.4$ and $\Come=273.8$}
 in the original Walecka fit); the comparison is a loose one, the two theories saturating at slightly different densities and bindings, but it suffices to establish that the dark sector contemplated here is strongly coupled in precisely the same sense as quantum hadrodynamics, where exchange and correlation corrections beyond the Hartree mean field are known to be appreciable, and the mean-field treatment is therefore adopted with the same scope and the same caveats \citep{1997IJMPE...6..515S}.
	Secondly, the scalar self-couplings of set~D respect the perturbative unitarity bounds on the mediator sector \citep{1977PhRvL..38..883L,1977PhRvD..16.1519L,2018EPJC...78..748G} by a comfortable margin, whilst reducing the incompressibility from $0.672$ to $0.334\,m_1$ and raising the effective mass from $0.54$ to $0.78\,m_1$, the same service that the Boguta--Bodmer terms render to nuclear matter, where they repair an incompressibility that would otherwise be embarrassingly stiff (the set-A value, $K=0.672\,m_1$, translates at the nucleon mass to $631\,\mathrm{MeV}$, against an empirical value near $240\,\mathrm{MeV}$) \citep{1977NuPhA.292..413B}.

	\begin{revisedtext}
	The nuclear analogy thus extends to the mechanism, to the field content, and even to the magnitudes of the couplings, but it should not be pressed further than the physics allows.
	What the two systems share is the competition between a scalar attraction whose source is quenched at high density and a vector repulsion whose source is not, which produces a minimum of $E/A$ at finite density with a binding of order a per cent of the particle mass (the nuclear value, $16\,\mathrm{MeV}$ per nucleon, is $1.7$ per cent of the nucleon mass).
	What separates them is everything that ordinary nuclear matter carries in addition: the dark fermions bear neither electric charge nor isospin, so there is no Coulomb repulsion, no $\rho$-meson channel, no symmetry energy, and no $\beta$-equilibrium with leptons \citep{1997csnp.book.....G}.
	The consequence is decisive for stellar structure.
	Nuclear matter saturates only when it is nearly symmetric in isospin; the neutron-rich matter of a neutron star, penalised by the symmetry energy, is unbound and must be confined by gravity, so neutron stars end in a crust and a diffuse envelope.
	The dark matter considered here suffers no such penalty and is self-bound at any size, which is why its stars, unlike neutron stars, terminate at a sharp surface.
	The fiducial matter is, moreover, stiffer than its nuclear counterpart, and its particle mass $m_1$ is a free parameter rather than a measured one.
	\end{revisedtext}

	\subsection{The equation of state and the saturation window}
	\label{sec:eos_results}

	\begin{figure}
	\centering
	\includegraphics[width=\columnwidth]{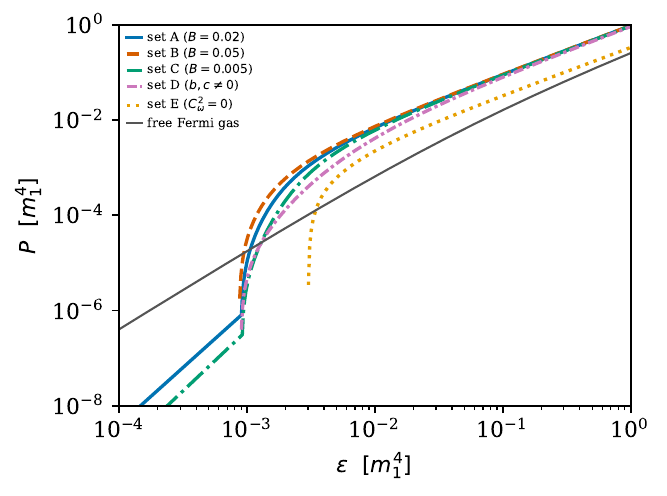}
	\caption{Pressure against energy density for the parameter sets of Table~\ref{tab:sets} and for the free Fermi gas (thin grey). The self-bound sets terminate at $P=0$ with finite energy density (left edge of each curve), the signature of a sharp stellar surface; the faint disconnected segments at the lower left are the dilute fragments of sets A and C, where kinetic pressure again exceeds the attraction. At high density the vector channel dominates the saturating sets, with $P\to\varepsilon$ from below and the sound speed approaching, without exceeding, the causal limit, whilst the scalar-only set and the free gas tend instead to the ultrarelativistic $P=\varepsilon/3$.}
	\label{fig:eos}
	\end{figure}

	Figure~\ref{fig:eos} shows the resulting barotropes.
	Their two ends tell the story: at low pressure each saturating set terminates at finite energy density, $\varepsilon_{\mathrm{surf}}=(1-B)\,m_1\nsat\simeq m_1\nsat$, the bulk matter simply ending where $P=0$ (the scalar-only set ends likewise, at its own far deeper coexistence point near $3.3\,m_1\nsat$), whilst at high density the vector condensate contributes $\tfrac{1}{2}\momega^2\omega_0^2\propto n^2$ to both $\varepsilon$ and $P$, driving the stiffness upward until the sound speed approaches the causal limit from below; in the fiducial set the squared sound speed reaches $c_s^2=0.81$ at the centre of the maximum-mass star and $0.99$ at the highest densities tabulated, so causality is respected throughout, as it must be in a Lagrangian-based equation of state \citep{1997csnp.book.....G}.
	The contrast with the scalar-only set is stark at every density: lacking the $n^2$ repulsion, set~E remains soft until the effective mass has been extinguished and the kinetic pressure of what are by then nearly massless fermions takes over.

	The saturation conditions cannot be met for arbitrarily strong vector coupling, and the limit is worth recording because it bounds the entire family of theories.
	Along the one-parameter family defined by $P(\nsat)=0$ at $b=c=0$, with $\Csig$ refitted at each $\Come$, the binding fraction passes through zero at $\Come\simeq 231$ (below which the minimum sits at positive energy and the matter is merely metastable) and the saturation condition loses all solutions at
	\begin{equation}
		\Comemax \simeq 512 \qquad (x_{\mathrm{sat}}=0.30)\,,
		\label{eq:Cv_max}
	\end{equation}
	beyond which the repulsion overwhelms the maximal scalar attraction, the latter being capped by the vanishing of $m^*$, and no zero of the pressure exists at the prescribed density.
	A dark sector with $\Come>\Comemax$ supports no self-bound matter at this density, however large its scalar coupling, a statement with the pleasing property of being falsifiable at the level of the Lagrangian.

	\section{Stellar structure and the self-consistent maximum mass}
	\label{sec:stars}

	\subsection{Density profiles and the self-bound surface}
	\label{sec:profiles}

	\begin{figure}
	\centering
	\includegraphics[width=\columnwidth]{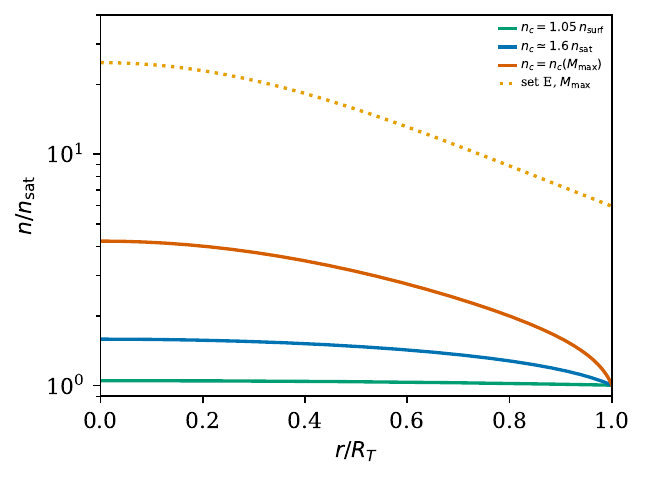}
	\caption{Radial density profiles for set~A at three central densities (solid curves): just above the surface density, at $n_c\simeq 1.6\,\nsat$, and at the central density of the maximum-mass configuration ($n_c\simeq 4.2\,\nsat$). All terminate at the surface with $n=\nsat$, the bulk coexistence density. The dotted curve shows the scalar-only set~E at its own maximum mass, whose surface density ($5.9\,\nsat$) and central density ($\simeq 25\,\nsat$) are both far higher.}
	\label{fig:profiles}
	\end{figure}

	Figure~\ref{fig:profiles} displays the density profiles of fiducial-set stars across the equilibrium sequence.
	The qualitative resemblance to strange quark stars \citep{1984PhRvD..30..272W,1986ApJ...310..261A,1986A&A...160..121H} is no accident: self-bound matter produces, at low central density, nearly uniform configurations held together by their own binding with gravity an afterthought, and, near the maximum mass, centrally concentrated configurations in which gravity has reasserted itself; in every case the density terminates abruptly at the surface value $\nsat$ rather than declining through a diffuse atmosphere.
	The lightest members of the sequence are not stars in any gravitational sense but macroscopic dark nuclei, the degenerate descendants of the bound states of \citet{2014PhRvD..90e5030W} and \citet{2015JHEP...02..023W}, and the sequence ascends continuously from these nuggets, through planetary and stellar masses, to the maximum mass; where exactly a nugget becomes a star is a question of taste rather than physics, the equation~\eqref{eq:closure} being indifferent to the distinction.

	\begin{revisedtext}
	The sharpness of the surface has a simple origin.
	Hydrostatic equilibrium requires the pressure to fall to zero at the edge of the star, and for self-bound matter this happens at the finite density $\nsat$, because uniform matter at $\nsat$ is already in equilibrium with the vacuum; any lower density would carry negative pressure and cannot form part of a static configuration.
	The density therefore drops from $\nsat$ to zero within the surface layer of Section~\ref{sec:bohm_surface}, a few Compton wavelengths thick.
	In an ordinary fermion star, built of free or purely repulsive fermions, the pressure vanishes only with the density itself; the outer layers are then non-relativistic polytropes of index $3/2$, and the density falls continuously, as $n\propto(R_T-r)^{3/2}$, over a sizeable fraction of the stellar radius.
	\end{revisedtext}

	\begin{revisedtext}
	The comparison with strange quark stars rests on the equation of state as well as on the profiles.
	In the MIT bag model the pressure of strange quark matter is $P=(\varepsilon-4B_{\rm bag})/3$, with $B_{\rm bag}$ the bag constant, so that the pressure vanishes at the finite energy density $4B_{\rm bag}$ \citep{1986ApJ...310..261A,1986A&A...160..121H}; the $\sigma$--$\omega$ barotropes of Fig.~\ref{fig:eos} share this property, terminating at $\varepsilon_{\mathrm{surf}}\simeq m_1\nsat$.
	The two families therefore display the same surface phenomenology: a density discontinuity on stellar scales, the $M\propto R^3$ rise of nearly uniform low-mass configurations, a maximum mass scaling as $\varepsilon_{\mathrm{surf}}^{-1/2}$, and the surface correction to the Love number of Appendix~\ref{app:kink}.
	They differ in two respects.
	The bag-model sound speed is fixed at $c_s^2=1/3$, whereas the vector channel drives it towards unity ($c_s^2=0.81$ at the centre of the fiducial maximum-mass star), so the dark matter is markedly stiffer at high density; and the quark surface carries an electrosphere, held by the Coulomb field of the quark charge \citep{1986ApJ...310..261A}, which has no counterpart in a dark sector without electromagnetic charge.
	\end{revisedtext}

	\subsection{Mass--radius relations and the maximum-mass map}
	\label{sec:MR}

	\begin{figure}
	\centering
	\includegraphics[width=\columnwidth]{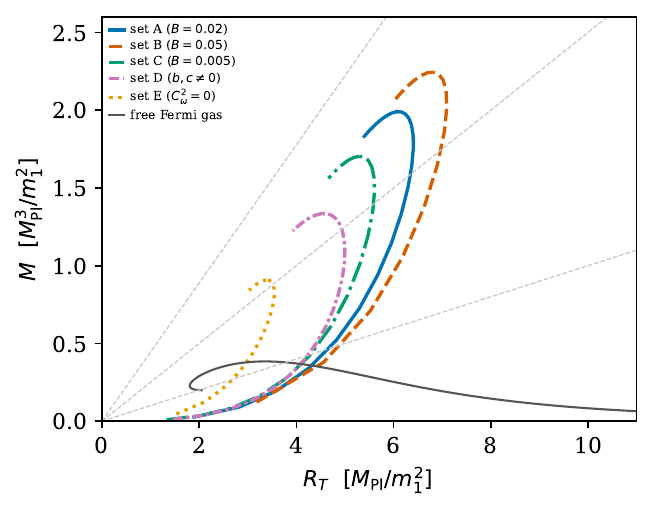}
	\caption{Mass--radius relations for the parameter sets of Table~\ref{tab:sets} and the free Fermi gas (grey). Solid portions denote the stable branch, the faded continuation a short stretch of the unstable branch beyond the maximum; thin grey dashed diagonals mark constant compactness $\kappa=0.1$, $0.25$, and the Buchdahl bound $4/9$. The self-bound sets display the $M\propto R^3$ rise characteristic of strange stars at low mass, in contrast with the inverse mass--radius trend of the gravitationally bound free gas.}
	\label{fig:MR}
	\end{figure}

	\begin{table}
	\centering
	\caption{Properties of the maximum-mass configuration for each parameter set: mass, radius, surface compactness $\kappa=GM/R_T$, central density in units of $\nsat$, and dimensionless tidal deformability $\Lambda$ at the maximum mass. Masses and radii are in the units of Eqs.~\eqref{eq:mass_unit} and \eqref{eq:length_unit}, respectively; the free-gas row is the Oppenheimer--Volkoff benchmark.}
	\label{tab:stars}
	\begin{tabular}{@{}lccccc@{}}
		\toprule
		Set & $M_{\rm max}\,[\Mpl^3/m_1^2]$ & $R_T\,[\Mpl/m_1^2]$ & $\kappa$ & $n_c/\nsat$ & $\Lambda$ \\
		\midrule
		A & $1.991$ & $6.07$ & $0.328$ & $4.2$ & $3.8$ \\
		B & $2.245$ & $6.82$ & $0.329$ & $3.3$ & $4.0$ \\
		C & $1.704$ & $5.32$ & $0.320$ & $5.4$ & $4.6$ \\
		D & $1.336$ & $4.55$ & $0.293$ & $8.0$ & $8.4$ \\
		E & $0.908$ & $3.38$ & $0.269$ & $24.9$ & $23.5$ \\
		free gas & $0.384$ & $3.36$ & $0.114$ & -- & $1.3\times 10^{3}$ \\
		\bottomrule
	\end{tabular}
	\end{table}

	\rev{Throughout, stability against radial perturbations is assessed by the turning-point criterion: along a sequence of fixed composition, configurations with $\mathrm{d}M/\mathrm{d}\varepsilon_c>0$ are stable and those with $\mathrm{d}M/\mathrm{d}\varepsilon_c<0$ are unstable, the squared frequency of the fundamental radial mode changing sign at each extremum of $M(\varepsilon_c)$ \citep{1983bhwd.book.....S}; the maximum-mass configuration therefore terminates the stable branch, and the same criterion is applied to each branch of Fig.~\ref{fig:branches}.}
	The mass--radius relations (Fig.~\ref{fig:MR}) and the maximum-mass properties (Table~\ref{tab:stars}) quantify the stiffening.
	At fixed particle mass, the fiducial $\sigma$--$\omega$ star reaches $M_{\rm max}=1.99\,\Mpl^3/m_1^2$, more than double the scalar-only value and five times the free-gas benchmark, at a radius nearly twice that of the scalar-only star; the deep-binding set~B reaches $2.25\,\Mpl^3/m_1^2$.
	The maximum compactness, $\kappa\simeq 0.33$ across sets A--C, approaches but respects the Buchdahl bound of $4/9$ \citep{1959PhRv..116.1027B} and comfortably exceeds that of a canonical neutron star ($\kappa\simeq 0.18$).

	\begin{figure}
	\centering
	\includegraphics[width=\columnwidth]{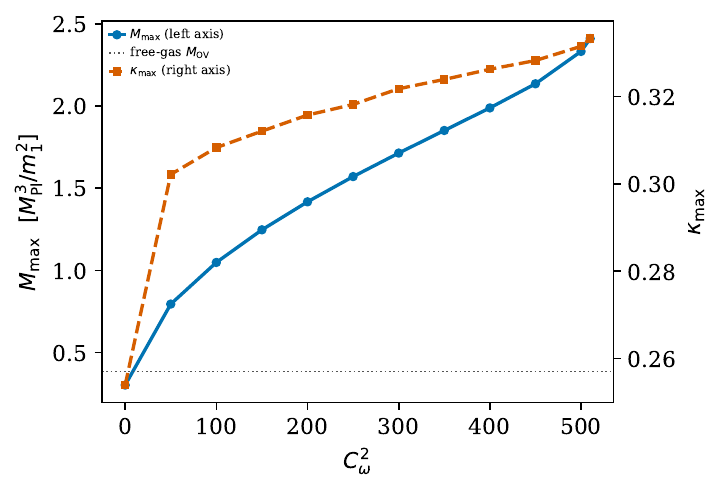}
	\caption{The self-consistent maximum mass (circles, left axis) and the corresponding surface compactness (squares, right axis) along the saturation constraint $P(\nsat)=0$, as functions of the vector coupling $\Come$ with $\Csig$ refitted at each point; the dotted line marks the free-gas Oppenheimer--Volkoff mass. The sequence terminates at $\Comemax\simeq 512$, Eq.~\eqref{eq:Cv_max}, where saturation ceases to exist; the last computed point, at $\Come=510$, reaches $M_{\rm max}=2.41\,\Mpl^3/m_1^2$.}
	\label{fig:Mmax_map}
	\end{figure}

	Figure~\ref{fig:Mmax_map} assembles the central quantitative result of this paper: the maximum mass as an explicit function of the vector coupling along the saturation constraint.
	$M_{\rm max}$ rises monotonically from $0.30\,\Mpl^3/m_1^2$ at $\Come=0$ (a metastable configuration, its energy minimum lying slightly above zero binding) through the free-gas value near $\Come\simeq 4$, to $2.41\,\Mpl^3/m_1^2$ at the edge of the saturation window, with the compactness climbing in parallel from $0.25$ to $0.33$.
	\rev{The eightfold rise has a transparent origin. Along the saturation constraint the surface energy density is pinned at $\varepsilon_{\mathrm{surf}}\simeq m_1\nsat$, so the mass scale set by the surface is common to the whole sequence, and the growth of $M_{\rm max}$ reflects the stiffness of the interior alone. That stiffness is governed by the vector coupling: by Eqs.~\eqref{eq:omega_alg} and \eqref{eq:CsCv} the vector condensate contributes $\tfrac{1}{2}\momega^2\omega_0^2=\tfrac{1}{2}(\Come/m_1^2)\,n^2$ to both $\varepsilon$ and $P$, so increasing $\Come$ drives the high-density matter towards the causal limit $P\to\varepsilon$, whilst $\Csig$, refitted to hold the saturation point fixed, merely follows. $\Come$ is thus the controlling parameter, and the stars grow in mass and radius as the equation of state approaches the stiffest form that causality permits.}
	The trend is the dark-sector incarnation of a result familiar from interacting fermion stars \citep{2006PhRvD..74f3003N,2015PhRvD..92f3526K}, but with a feature those treatments lack: because saturation caps the vector coupling at $\Comemax$, the theory possesses an absolute ceiling,
	\begin{equation}
		M_{\rm max} \;\lesssim\; 2.4\,\frac{\Mpl^3}{m_1^2}
		= 3.9\,\Bigl(\frac{m_1}{\mathrm{GeV}}\Bigr)^{\!-2} M_\odot\,,
		\label{eq:Mmax_ceiling}
	\end{equation}
	valid across the entire family of self-bound $\sigma$--$\omega$ theories saturating at $x_{\mathrm{sat}}=0.30$\rev{ with $b=c=0$. It is not a property of the fiducial set, whose own maximum is $3.25\,\Msun$ at $m_1=1\,\mathrm{GeV}$, but the supremum of $M_{\rm max}$ over every vector coupling compatible with saturation on that slice; the soft set~D, whose self-couplings lower the incompressibility at fixed binding, lies below it (Table~\ref{tab:stars})}.
	The coefficient is tied to that fiducial slice: since the maximum mass of self-bound matter scales as the inverse square root of its surface energy density, in the manner familiar from strange quark stars, it is expected to rescale approximately as $x_{\mathrm{sat}}^{-3/2}$, whilst the existence of a coupling cap, which descends from the vanishing of the effective mass rather than from any property of the chosen density, survives the choice; it is the structure of the bound, not its coefficient, that constitutes the prediction.
	A measured mass above this ceiling, at a particle mass $m_1$ fixed by the broader dark-matter context, would falsify the framework outright; conversely a population of dark compact objects clustering below it would constrain the coupling pair $(\Csig,\Come)$ through Fig.~\ref{fig:Mmax_map} directly.
	It is the conversion of Lagrangian parameters into a maximum mass, rather than any single number, that distinguishes the present construction from its scalar-only counterpart.

	\subsection{Branch structure}
	\label{sec:branches}

	\begin{figure}
	\centering
	\includegraphics[width=\columnwidth]{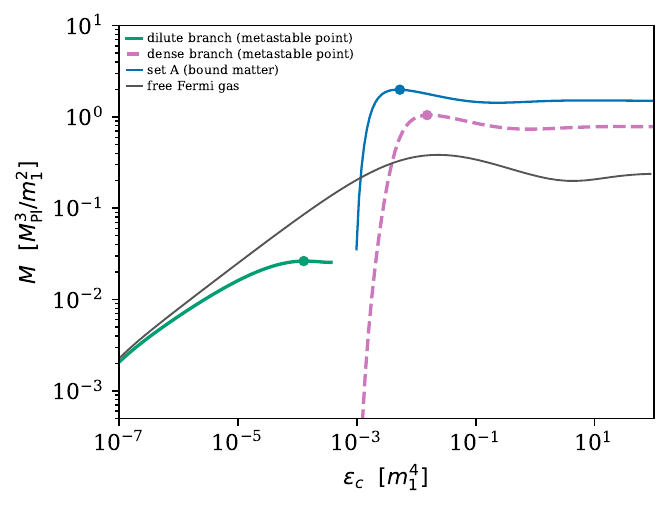}
	\caption{Stellar mass against central energy density for the metastable parameter point ($\Come=100$, $\Csig=154.1$, saturation at positive energy): the dilute gravitationally bound branch (green) terminates at the mechanical-stability boundary of uniform matter, and the dense self-bound branch (purple dashed) begins at the coexistence density; filled circles mark the maximum-mass configurations. Set~A (blue) and the free Fermi gas (grey) are shown for comparison. No equilibria exist in the gap, where uniform matter has negative pressure.}
	\label{fig:branches}
	\end{figure}

	Whether the theory supports one family of stars or two is decided by the sign of the binding.
	For strictly bound matter ($B>0$, sets A--D) the dilute uniform phase is thermodynamically unstable against fragmentation into nuggets, and the equilibrium sequence is the single self-bound family already described.
	For metastable matter ($0<\Come\lesssim 231$ along the $b=c=0$ saturation constraint, where the minimum of $E/A$ sits at positive energy), the low-density phase survives, and with it a second family: dilute, gravitationally bound configurations supported by degeneracy pressure, the direct descendants of the soft stars of the scalar-only theory.
	\rev{Metastability is here a statement about the bulk matter rather than about the stars. The saturation point remains a local minimum of $E/A$, with $P=0$ and positive incompressibility, so uniform matter at $\nsat$ is mechanically stable; but the minimum lies above $m_1$, so the dense phase is not bound with respect to free fermions at rest, and its hydrostatic configurations are local rather than global minima of the energy, which in the absence of sufficient gravitational binding could be lowered by dispersal, a process the hydrostatic treatment does not follow. Where the gravitational redshift is large enough that $\mu_s\,e^{\nu(R_T)}<m_1$, the star is bound globally as well. Each family is, in addition, separately stable against radial perturbations by the turning-point criterion of Section~\ref{sec:MR}; which family a given dark object occupies is decided by its formation history, a question outside the scope of this paper.}
	Figure~\ref{fig:branches} displays the resulting branch structure at the representative point $\Come=100$.
	The dilute branch ascends to a modest maximum, $M=0.026\,\Mpl^3/m_1^2$ at $\kappa\simeq 3\times 10^{-3}$, and terminates where the central density reaches the mechanical-stability boundary of the uniform phase; after a gap in central density within which no static configuration exists, the dense branch rises from the coexistence density to $M_{\rm max}=1.05\,\Mpl^3/m_1^2$ at $\kappa=0.31$.
	The two families differ in maximum mass by a factor of forty and in tidal deformability by more than ten orders of magnitude, so they could scarcely be confused observationally; the structure is the dark analogue of the white-dwarf and neutron-star branches, with the spinodal of the dark fluid playing the role of neutronisation.
	The visible families carry no such contrast: the white-dwarf and neutron-star maxima nearly coincide, both being anchored to the Chandrasekhar combination $\Mpl^3/m_B^2$, with $m_B$ the baryon mass, whereas the factor of forty is possible here precisely because the dense family answers to the couplings rather than to degeneracy.
	Finally, I note for completeness that within each individual barotrope explored here the stable branch is single, with no twin-star wiggle: producing twins would require a phase-transition-like softening interior to the dense phase, which the present field content does not supply.

	\section{The Bohm surface layer and two-species structure}
	\label{sec:surface}

	\subsection{The quantum surface of a self-bound star}
	\label{sec:bohm_surface}

\begin{figure}
	\centering
	\includegraphics[width=\columnwidth]{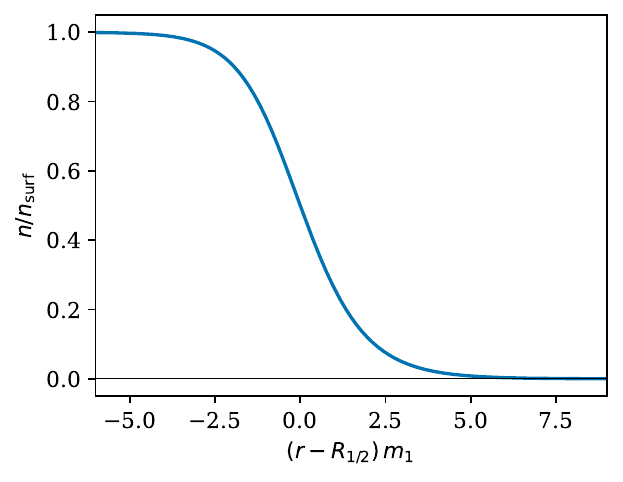}
	\caption{The Bohm surface kink of the fiducial set: density across the stellar surface in units of the bulk coexistence density, against proper distance in units of $1/m_1$, centred on the half-density point. The $10$--$90$ per cent thickness is $4.1/m_1$; for $m_1=1\,\mathrm{GeV}$ the layer is less than a femtometre thick on a star of $14.6\,\mathrm{km}$.}
	\label{fig:kink}
	\end{figure}

	In the scalar-only theory the Bohm term shapes the entire dilute envelope; saturation changes its station in life, confining it to a thin layer at the sharp surface, where the density gradient, and hence the quantum potential of Eq.~\eqref{eq:bohm}, is largest.
	Because the layer is thin compared with the stellar radius, the surface may be treated as locally planar.
	Writing $\psi=\sqrt{n}$ and balancing the Bohm term against the bulk thermodynamics at fixed surface chemical potential $\mu_s$, the closure~\eqref{eq:closure} reduces to a one-dimensional equation whose first integral,
	\begin{equation}
		\frac{\lamB}{2m_s^*}\left(\frac{\mathrm{d}\psi}{\mathrm{d}z}\right)^{\!2} = w(\psi^2)\,,\qquad
		w(n)\equiv\varepsilon(n)-\mu_s\,n\,,
		\label{eq:kink}
	\end{equation}
	connects the bulk density $n=\nsat$ (where $w=0$ by the coexistence condition $P(\nsat)=0$) to the vacuum (where $w\to 0$ likewise); $m_s^*$ is the effective mass at the surface, and the derivation, together with the surface-tension quadrature, is collected in Appendix~\ref{app:kink}.
	The profile is the quantum-hydrodynamic kink shown in Fig.~\ref{fig:kink}.
	For the fiducial set the $10$--$90$ per cent thickness is $t=4.1/m_1$ and the surface tension, denoted $\sigma_s$ to keep it distinct from the mediator field, is $2.42\times 10^{-5}\,m_1^3$; the deeper the binding the thinner and tenser the surface ($t=3.1/m_1$, $\sigma_s=4.18\times 10^{-5}\,m_1^3$ for set~B), the shallower the binding the broader and weaker ($t=6.1/m_1$, $\sigma_s=1.31\times 10^{-5}\,m_1^3$ for set~C), as one would expect of any interface whose width is set by the geometric mean of the gradient stiffness and the well depth.
	\rev{These numbers follow from the first integral~\eqref{eq:kink} alone, with no stellar integration: the profile $z(\psi)$ is obtained by the quadrature of Eq.~\eqref{eq:app_profile}, the thickness $t$ is the distance between the points at which $n$ falls to $90$ and $10$ per cent of $\nsat$, and $\sigma_s$ is the integral of Eq.~\eqref{eq:app_sigma}; dimensionally, $t\sim\sqrt{\lamB\,\nsat/(2m_s^*\bar w)}$, with $\bar w$ a typical value of $w$ between its two zeros, which for the fiducial set amounts to a few Compton wavelengths.}
	One should be candid about the accounting: meson-field gradients, neglected here in the local-density spirit, contribute surface energies of the same order, as detailed studies of dark nugget surfaces have shown \citep{2017PhRvD..96i6012G,2018PhRvD..98i6001G}, so $\sigma_s$ as computed isolates the specifically quantum-kinetic share of the surface energy, which is the novel ingredient of the present framework.
	A further candour concerns wavenumber: the inverse width of the kink is comparable to the surface Fermi momentum itself, so the gradient expansion underlying Eq.~\eqref{eq:bohm} operates here at the edge of its domain of validity, and the surface quantities are to be read as controlled estimates rather than precision results; the alternative prescription in which the gradient inertia is carried by the Landau mass $\sqrt{p_F^2+m^{*2}}$ rather than by $m^*$ shifts the thickness and tension by less than ten per cent, which is some comfort.

	Two consequences follow, one consequential and one reassuring.
	The consequential one concerns the smallest objects in the family.
	In the liquid-drop spirit, a nugget of $N$ particles has energy $-B\,m_1 N + a_S N^{2/3}$ with the surface coefficient $a_S=4\pi\sigma_s\,[3/(4\pi\nsat)]^{2/3}$, whence stability against evaporation requires $N>N_{\rm min}=(a_S/Bm_1)^3$; for the fiducial and deep sets $N_{\rm min}$ is below unity and every nugget is bound, but for the shallow-binding set~C one finds $a_S=6.7\times 10^{-3}\,m_1$ and $N_{\rm min}\simeq 2.4$, so the Bohm surface energy forbids the smallest dark nuclei outright, in qualitative agreement with the synthesis bottlenecks discussed for dark nucleosynthesis \citep{2015JHEP...06..011H,2018PhRvD..98i6001G}; the cubic sensitivity of $N_{\rm min}$ to the surface tension counsels reading the threshold as indicative rather than sharp.
	The reassuring one concerns the stars: the surface correction to the gravitational mass is $\Delta M/M = 4\pi R_T^2\sigma_s/M\simeq 5.6\times 10^{-3}\,(m_1/\Mpl)$ at the fiducial maximum mass, which for $m_1=1\,\mathrm{GeV}$ evaluates to $4.6\times 10^{-22}$, a number I am content to describe as negligible.
	\rev{The estimate is analytic rather than numerical: it is the surface energy $4\pi R_T^2\sigma_s$ of the computed configuration divided by its mass, and not the difference between two TOV integrations, which double-precision arithmetic could not resolve at the level of $10^{-22}$. Its accuracy is therefore that of $\sigma_s$, of order ten per cent from the treatment of $m_s^*$ and comparable again from the gradient expansion and the omitted meson-field gradients; an error of even an order of magnitude would leave the conclusion unchanged.}
	\rev{The physical reason for the smallness is geometric. The Bohm term acts only where the density varies on the Compton scale $1/m_1$, that is, within the surface layer, so its contribution grows with the surface, as $N^{2/3}$, whilst the bulk binding grows as $N$; their ratio falls as $N^{-1/3}$, or equivalently as $t/R_T$. For a nugget of a few particles the two are comparable, and the surface term decides whether the object exists at all; for a star, with $R_T\sim\Mpl/m_1^2$ and $t\sim 1/m_1$, the ratio is proportional to $m_1/\Mpl$, about $8\times 10^{-20}$ for $m_1=1\,\mathrm{GeV}$, and the quantum pressure is irrelevant to the stellar structure.}
	The Bohm term therefore polices the small-$N$ end of the population whilst leaving the stellar observables untouched, which is precisely the division of labour one wants from a surface term.

	\subsection{Two-species configurations}
	\label{sec:twospecies}

	\begin{figure}
	\centering
	\includegraphics[width=\columnwidth]{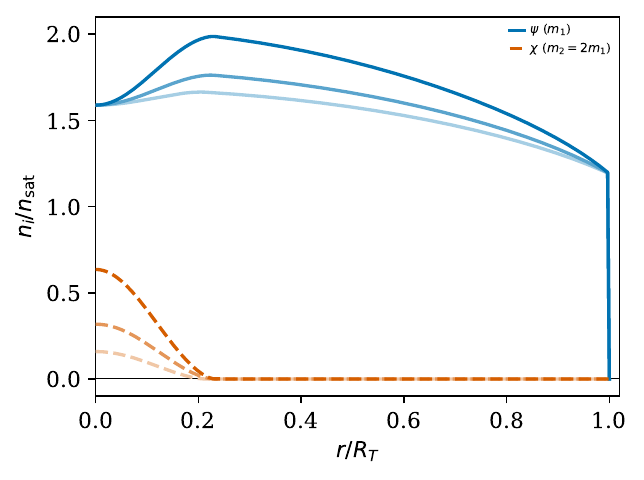}
	\caption{Two-species configurations with the fiducial couplings, a heavier minority species ($m_2=2m_1$, equal couplings), and central density $n_1(0)=1.6\,\nsat$: number densities of $\psi$ (solid) and $\chi$ (dashed) against radius for central number-density fractions $n_2(0)/n_1(0)=0.1$, $0.2$, and $0.4$, with line intensity increasing with the admixture. The heavy species is confined in every case to a core inside $r\simeq0.23\,R_T$, the light-species density responds visibly to the edge of the $\chi$ core, and the exterior is a pure-$\psi$ body terminating at the common self-bound surface. The fiducial $0.2$ configuration has $M=0.823\,\Mpl^3/m_1^2$, $R_T=5.31\,\Mpl/m_1^2$, and $\kappa=0.155$.}
	\label{fig:twospecies}
	\end{figure}

	The two-species machinery of Section~\ref{sec:tov} is exercised in Fig.~\ref{fig:twospecies} for a configuration in which a heavier minority species $\chi$ ($m_2=2m_1$, with equal mediator couplings) is admixed at central number-density fractions $n_2(0)/n_1(0)=0.1$, $0.2$, and $0.4$.
	The double closure~\eqref{eq:closure} segregates the species cleanly: the heavier fermions, more strongly bound gravitationally per particle yet identically repelled by the shared vector field, are confined to a core occupying roughly the inner fifth of the stellar radius, beyond which the star is a pure-$\psi$ body terminating at the usual self-bound surface; the kink in the $\psi$ profile at the core boundary records the discontinuity in the sources of the shared meson fields. The confinement is robust rather than tuned: as the admixture is quadrupled the fractional core radius barely moves, from $0.21\,R_T$ to $0.23\,R_T$, whilst the stellar mass grows from $0.654$ to $1.154\,\Mpl^3/m_1^2$, the additional binding deepening the $\psi$ profile rather than spreading the core.
	The configurations shown satisfy the turning-point criterion along sequences of fixed composition; a complete stability analysis for matter carrying two separately conserved charges would require the composition to be varied as well, and is not attempted here.
	The core edge is, moreover, a quantum boundary in its own right.
	Near $r_c\simeq0.23\,R_T$, the internal radius at which the $\chi$ Fermi momentum is exhausted whilst the star continues as a pure-$\psi$ body out to the true surface at $R_T$, the closure~\eqref{eq:closure} pins $\mu_2\,e^{\nu}$ to a constant whilst the metric and meson fields, sourced predominantly by the ambient $\psi$ fluid, remain smooth across the edge.
	The local Fermi momentum therefore vanishes as $p_{F,2}^2\propto(r_c-r)$, the Thomas--Fermi density as $n_2\propto p_{F,2}^{3}\propto(r_c-r)^{3/2}$, and the gradient term of Eq.~\eqref{eq:bohm} consequently grows as the inverse square of the distance to the edge, necessarily dominating within a collar of Compton width.
	The collar is an internal counterpart of the surface layer of Section~\ref{sec:bohm_surface}, though of opposite character: the core edge is soft, the density vanishing continuously, whereas the self-bound surface falls from the bulk density to zero within a few Compton wavelengths, abrupt on every stellar scale.
	Its resolution is the kink analysis of Appendix~\ref{app:kink} transplanted to the core boundary, with the ambient $\psi$ fluid supplying the background fields, and the corrections it returns to the star are suppressed by the same powers that render the surface contribution of order $10^{-22}$, so the configurations shown are unaffected.
	The structure is the self-bound counterpart of the core--halo morphology familiar from dark-matter-admixed neutron stars \citep{2017PhRvD..96h3004P,2020PhRvD.102f3028I,2024PhR..1052....1B}, realised here within a single dark sector; since both species couple to the same $\omega$, the vector repulsion enforces a common surface chemistry, and no dilute $\chi$ halo forms in the parameter range explored.

	\section{Observational signatures}
	\label{sec:observations}

	\subsection{Physical scales}
	\label{sec:scales}

	\begin{table}
	\centering
	\caption{Physical observables of the fiducial (set~A) maximum-mass configuration as functions of the dark fermion mass, from Eqs.~\eqref{eq:length_unit} and \eqref{eq:mass_unit}: maximum mass, radius, and contact gravitational-wave frequency $f_{\rm GW}=\pi^{-1}\sqrt{GM_{\rm tot}/(2R_T)^3}$ for the equal-mass binary at touching separation.}
	\label{tab:physical}
	\begin{tabular}{@{}cccc@{}}
		\toprule
		$m_1$ [GeV] & $M_{\rm max}$ [$\Msun$] & $R_T$ [km] & $f_{\rm GW}$ [Hz] \\
		\midrule
		$0.1$ & $325$   & $1463$  & $19$ \\
		$0.3$ & $36.1$  & $163$   & $168$ \\
		$1.0$ & $3.25$  & $14.6$  & $1.9\times 10^{3}$ \\
		$3.0$ & $0.36$  & $1.6$   & $1.7\times 10^{4}$ \\
		\bottomrule
	\end{tabular}
	\end{table}

	The dimensionless results acquire physical content once $m_1$ is specified, and Table~\ref{tab:physical} performs the conversion for the fiducial set.
	The window of greatest observational interest is $m_1\sim 0.3$--$3\,\mathrm{GeV}$, where the maximum mass spans the subsolar regime through the lower mass gap: at $m_1=1\,\mathrm{GeV}$ the family terminates at $3.25\,\Msun$ and $14.6\,\mathrm{km}$, numbers that any practising astronomer will recognise as provocatively neutron-star-like, whilst at $m_1=3\,\mathrm{GeV}$ the entire population is subsolar, a regime in which any compact-object detection would be unarguably exotic.
	Lighter fermions push the family to intermediate-mass scales accessible at lower frequencies.
	It deserves emphasis that the saturation regime studied here occupies the GeV window naturally suggested by asymmetric dark matter \citep{2014PhRvD..90e5030W,2018PhR...730....1T}, and is thereby complementary to the ultralight regime in which the Bohm potential is globally relevant \citep{LopesQuantumStarsNewtonian2026}, the two regimes bracketing many orders of magnitude in particle mass with the same equilibrium closure, there in its Newtonian and here in its relativistic form.

	\subsection{Tidal deformability as a three-way discriminant}
	\label{sec:tidal}

	\begin{figure*}
	\centering
	\includegraphics[width=0.92\textwidth]{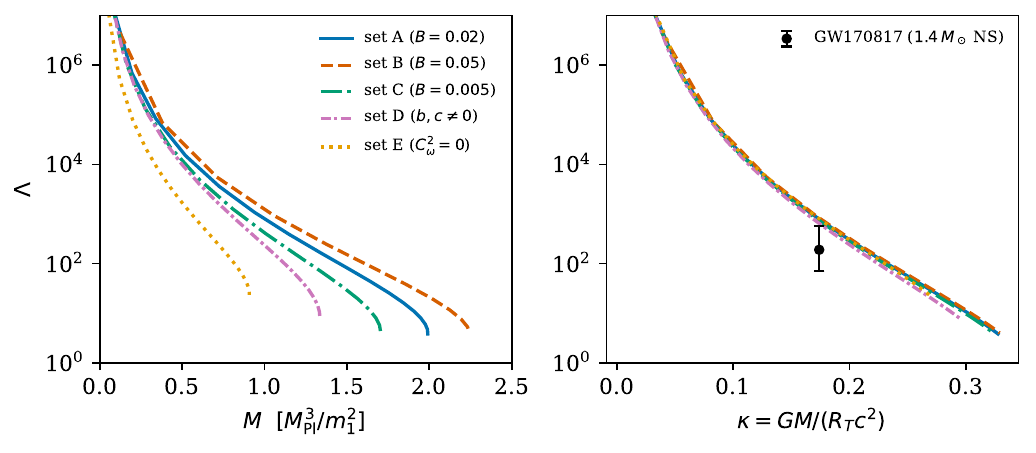}
	\caption{Dimensionless tidal deformability of the stable branch for the parameter sets of Table~\ref{tab:sets}, against mass (left) and against compactness (right). At fixed mass the vector channel raises $\Lambda$ by one to two orders of magnitude relative to the scalar-only set; at fixed compactness the curves collapse onto a quasi-universal band. The black point with error bars marks the GW170817 measurement for a $1.4\,M_\odot$ neutron star, $\Lambda=190^{+390}_{-120}$ \citep{2018PhRvL.121p1101A}, at the corresponding compactness $\kappa\simeq 0.174$.}
	\label{fig:tidal}
	\end{figure*}

	The dimensionless tidal deformability $\Lambda=\tfrac{2}{3}k_2\,\kappa^{-5}$, with $k_2$ the quadrupolar Love number computed in the standard manner \citep{2008ApJ...677.1216H,2008PhRvD..77b1502F} including the surface-discontinuity correction appropriate to self-bound stars (Appendix~\ref{app:kink}), is displayed in Fig.~\ref{fig:tidal}, and it is here that the three-way discrimination promised in the Introduction becomes quantitative.

	Consider first the internal comparison, $\sigma$--$\omega$ against scalar-only, at fixed $m_1=1\,\mathrm{GeV}$.
	A $1.4\,\Msun$ member of the fiducial family has $R_T=13.3\,\mathrm{km}$, $\kappa=0.155$, and $\Lambda\simeq 1.6\times 10^{3}$; its scalar-only counterpart has $R_T=8.5\,\mathrm{km}$, $\kappa=0.242$, and $\Lambda\simeq 65$, smaller by a factor of twenty-five.
	The separation dwarfs the measurement uncertainty already achieved for GW170817, whose ninety per cent interval spans a few hundred in $\Lambda$, so a single inspiral of comparable quality decides between the two matter sectors, which is the practical meaning of the statement that the vector field stiffens the equation of state.
	\rev{The comparison is made at matched microscopic parameters in the sense the two theories allow: set~E shares the particle mass and the scalar coupling $\Csig$ of set~A and differs solely by the absence of the vector channel, so that its mass--radius relation (Fig.~\ref{fig:MR}) and deformabilities (Fig.~\ref{fig:tidal}) isolate the effect of $\Come$. A comparison at a common saturation point is not available at that scalar coupling, since without the vector field the matter saturates only near $5.9\,\nsat$ (Section~\ref{sec:satfit}); the nearest equivalent is the $\Come\to 0$ end of the saturation constraint in Fig.~\ref{fig:Mmax_map}, where $\Csig$ is refitted to place a metastable minimum at $\nsat$ and the maximum mass falls to $0.30\,\Mpl^3/m_1^2$, below even the free-gas value, so that either way of matching the parameters leads to the same conclusion.}

	Consider next the comparison with neutron stars.
	The fiducial $1.4\,\Msun$ dark star carries a deformability almost nine times the GW170817 central value of $190^{+390}_{-120}$ \citep{2018PhRvL.121p1101A} and well outside its ninety per cent interval, whilst its radius lies comfortably within the band NICER infers for neutron-star radii \citep{2021ApJ...918L..28M}\rev{, and its compactness, $\kappa=0.155$, lies correspondingly a little below the value $\kappa\simeq 0.17$ of a $1.4\,\Msun$ neutron star of radius $12\,\mathrm{km}$}; the discrimination is thus carried by $\Lambda$ rather than by the radius, and GW170817 already disfavours the fiducial couplings as a description of that particular source, which is exactly how Fig.~\ref{fig:Mmax_map} is intended to be used in reverse.
	At fixed compactness, by contrast, the right panel of Fig.~\ref{fig:tidal} shows all five sets collapsing onto a quasi-universal band, a behaviour reminiscent of the universal relations of neutron-star physics; at the GW170817 compactness the band spans $\Lambda\simeq 700$--$1000$, some four times the measured central value, so even this blunter comparison, which imports the equation-of-state-inferred neutron-star radius and is to that extent illustrative rather than model-independent, retains discriminating power.

	\begin{revisedtext}
	These statements require two qualifications.
	The first concerns the neutron stars: the comparison is with the equations of state that survive current constraints, the GW170817 interval already folding in the measurement uncertainty, and stiffer hadronic models, now disfavoured by that event, predict deformabilities well above its central value \citep{2017PhRvL.119p1101A,2018PhRvL.121p1101A}; the contrast quoted above is therefore with the neutron stars that nature appears to make rather than with every conceivable one.
	The second concerns the dark stars: the figures quoted hold at fixed $m_1$ and fixed couplings, and because masses and radii rescale as $m_1^{-2}$ at fixed compactness, a heavier fermion, or a softer coupling choice, places a $1.4\,\Msun$ dark star closer to its maximum mass, at higher compactness and smaller $\Lambda$, and could bring its deformability within the neutron-star interval.
	A single measurement of $(M,\Lambda)$ thus discriminates cleanly between hypotheses that fix these parameters, whereas excluding the family as a whole requires the slope of $\Lambda(M)$ across several events, or the comparison at fixed compactness of the right panel of Fig.~\ref{fig:tidal}.
	In practice the measured quantity is the binary deformability $\tilde\Lambda$, whose uncertainty depends on the signal-to-noise ratio, the mass ratio, and the spin priors, and the discriminating power of any one event must be judged with those uncertainties in view.
	\end{revisedtext}

	Consider finally the comparison with black holes, for which $\Lambda=0$ identically \citep{2019LRR....22....4C}.
	Every configuration in Table~\ref{tab:stars} retains a non-zero deformability, $\Lambda=3.8$--$23$ at the maximum mass, so a high signal-to-noise inspiral terminating in the mass gap could in principle separate a maximal $\sigma$--$\omega$ star from a low-mass black hole; the measurement is demanding, as $\Lambda$ of order a few lies at the edge of foreseeable sensitivity even for third-generation instruments \citep{2010CQGra..27s4002P,2019BAAS...51g..35R} and may well remain beyond them, but the obstacle is one of sensitivity rather than of principle.
	Mass-gap candidates of the kind already observed \citep{2020ApJ...896L..44A} are the natural hunting ground, since Eq.~\eqref{eq:Mmax_ceiling} places the fiducial family's terminus at $3.9\,(m_1/\mathrm{GeV})^{-2}\,\Msun$, squarely in the gap for $m_1\simeq 1$--$1.3\,\mathrm{GeV}$.
	Indeed, at $m_1=1\,\mathrm{GeV}$ the heavier self-bound families (sets A--C) cross the GW190814 mass band still carrying $\Lambda\simeq 12$--$130$; that event itself returned no tidal information, but a louder successor would separate the three hypotheses, dark star, neutron star, and black hole, in a single measurement.

	The contact frequencies of Table~\ref{tab:physical}, point estimates that neglect the tidal hydrodynamics of the final orbits, complete the picture: equal-mass binaries of the $m_1=1\,\mathrm{GeV}$ family merge near $1.4$--$1.9\,\mathrm{kHz}$ (the $1.4\,\Msun$ pair at $1.4\,\mathrm{kHz}$), within the band of the Einstein Telescope and Cosmic Explorer \citep{2010CQGra..27s4002P,2019BAAS...51g..35R}, whilst the subsolar $m_1=3\,\mathrm{GeV}$ family merges above $10\,\mathrm{kHz}$, observable only through its inspiral, and the intermediate-mass $m_1\lesssim 0.3\,\mathrm{GeV}$ families descend towards the deci-hertz regime.
	The absence of any electromagnetic counterpart, of tidal disruption flares, and of post-merger neutrino emission would accompany all of these, as for any dark-sector object \citep{2019LRR....22....4C,2024PhR..1052....1B}.

	\section{Conclusions}
	\label{sec:conclusions}

	I have constructed and solved the curved-spacetime $\sigma$--$\omega$ theory of quantum dark fermion stars: two fermion species coupled to a scalar mediator through the effective mass and to a neutral vector mediator through the conserved number current, with renormalisable scalar self-interactions, closed by the two-fluid TOV--Bohm system and the redshifted chemical potentials of Eq.~\eqref{eq:closure}, the species-wise force balance being derived in full from the field equations.

	Three quantitative results follow.
	First, scalar--vector competition produces dark-sector saturation, and with it self-bound stars: sharp surfaces at the coexistence density, the $M\propto R^3$ rise of strange quark stars at low mass, and macroscopic dark nuggets at the light end.
	Saturation exists only for $\Come\lesssim\Comemax\simeq 512$ at the fiducial saturation density, and along the saturation constraint the maximum mass climbs monotonically from $0.30$ to $2.41\,\Mpl^3/m_1^2$ with the sound speed everywhere causal, whence the ceiling of Eq.~\eqref{eq:Mmax_ceiling}, $M_{\rm max}\lesssim 3.9\,(m_1/\mathrm{GeV})^{-2}\,\Msun$, an upper bound written directly in the couplings: a single heavier compact object, at a particle mass fixed by the wider dark-matter context, falsifies the entire family.
	Secondly, the vector channel stiffens the equation of state: at $m_1=1\,\mathrm{GeV}$ the $1.4\,\Msun$ fiducial star has $R_T=13.3\,\mathrm{km}$ and $\Lambda\simeq 1.6\times 10^{3}$, against $8.5\,\mathrm{km}$ and $\Lambda\simeq 65$ for its scalar-only counterpart, a factor of twenty-five in the cleanest gravitational-wave observable; metastable couplings ($\Come\lesssim 231$ at $b=c=0$) add a second, dilute stable family, separated from the self-bound one by a forbidden band of central densities, with maximum masses differing by a factor of forty and deformabilities by more than ten orders of magnitude.
	Thirdly, the Bohm correction, compressed into a surface layer of thickness $\simeq 4/m_1$ and tension $2.4\times 10^{-5}\,m_1^3$, supplies the quantum-kinetic share of the surface energy, sets the minimum nugget number at shallow binding ($N_{\rm min}\simeq 2.4$), and perturbs stellar masses below $\Delta M/M\sim 10^{-21}$: the quantum term governs the smallest objects in the theory whilst leaving its stars to general relativity, a hierarchy of consequence and irrelevance delivered in a single stroke.

	The originality of the construction lies in the combination: a saturating two-mediator dark sector had not previously been solved in curved spacetime, nor closed species by species with the vector potential and the quantum potential carried on the same footing as the gravitational redshift, and the coupling-capped maximum mass that results has no counterpart amongst scalar-only or self-interacting fermion stars.
	The signatures fall within reach of the coming generation of detectors: equal-mass binaries of the $m_1=1\,\mathrm{GeV}$ family merge at $1.4$--$1.9\,\mathrm{kHz}$, inside the bands of the Einstein Telescope and Cosmic Explorer \citep{2010CQGra..27s4002P,2019BAAS...51g..35R}; the family terminus crosses the lower mass gap for $m_1\simeq 1$--$1.3\,\mathrm{GeV}$, where candidate objects already exist \citep{2020ApJ...896L..44A}; and the deformability contrasts are large enough for a single high signal-to-noise inspiral\rev{, at given particle mass and couplings,}
 to decide amongst the $\sigma$--$\omega$ star, its scalar-only counterpart, and a neutron star, with the black-hole hypothesis excluded the moment a non-zero deformability is resolved.
	One boundary of the survey deserves an equally sharp statement: with universal couplings, the shared $\omega$ enforces a common surface chemistry and forbids dilute haloes, so any core--halo morphology within a single dark sector requires differential vector coupling, $g_{\omega 2}\neq g_{\omega 1}$, the one qualitatively new phenomenon that the full seven-parameter family holds in reserve.
	Taken with the Newtonian instalment \citep{LopesQuantumStarsNewtonian2026}, the framework converts a four-parameter dark Lagrangian into mass--radius--deformability predictions at the precision of gravitational-wave measurement, and it is by those predictions that it stands to be judged.

	\begin{acknowledgments}
		I.~L. thanks the Funda\c{c}\~ao para a Ci\^encia e Tecnologia (FCT), Portugal, for the financial support to the Center for Astrophysics and Gravitation (CENTRA/IST/ULisboa) through grant No.\ UID/PRR/00099/2025 (\doi{10.54499/UID/PRR/00099/2025}) and grant No.\ UID/00099/2025 (\doi{10.54499/UID/00099/2025}).
	\end{acknowledgments}

	\bibliography{ILopes}

\begin{thebibliography}{45}%
\makeatletter
\providecommand \@ifxundefined [1]{%
 \@ifx{#1\undefined}
}%
\providecommand \@ifnum [1]{%
 \ifnum #1\expandafter \@firstoftwo
 \else \expandafter \@secondoftwo
 \fi
}%
\providecommand \@ifx [1]{%
 \ifx #1\expandafter \@firstoftwo
 \else \expandafter \@secondoftwo
 \fi
}%
\providecommand \natexlab [1]{#1}%
\providecommand \enquote  [1]{``#1''}%
\providecommand \bibnamefont  [1]{#1}%
\providecommand \bibfnamefont [1]{#1}%
\providecommand \citenamefont [1]{#1}%
\providecommand \href@noop [0]{\@secondoftwo}%
\providecommand \href [0]{\begingroup \@sanitize@url \@href}%
\providecommand \@href[1]{\@@startlink{#1}\@@href}%
\providecommand \@@href[1]{\endgroup#1\@@endlink}%
\providecommand \@sanitize@url [0]{\catcode `\\12\catcode `\$12\catcode
  `\&12\catcode `\#12\catcode `\^12\catcode `\_12\catcode `\%12\relax}%
\providecommand \@@startlink[1]{}%
\providecommand \@@endlink[0]{}%
\providecommand \url  [0]{\begingroup\@sanitize@url \@url }%
\providecommand \@url [1]{\endgroup\@href {#1}{\urlprefix }}%
\providecommand \urlprefix  [0]{URL }%
\providecommand \Eprint [0]{\href }%
\providecommand \doibase [0]{https://doi.org/}%
\providecommand \selectlanguage [0]{\@gobble}%
\providecommand \bibinfo  [0]{\@secondoftwo}%
\providecommand \bibfield  [0]{\@secondoftwo}%
\providecommand \translation [1]{[#1]}%
\providecommand \BibitemOpen [0]{}%
\providecommand \bibitemStop [0]{}%
\providecommand \bibitemNoStop [0]{.\EOS\space}%
\providecommand \EOS [0]{\spacefactor3000\relax}%
\providecommand \BibitemShut  [1]{\csname bibitem#1\endcsname}%
\let\auto@bib@innerbib\@empty
\bibitem [{\citenamefont {{Bramante}}\ and\ \citenamefont
  {{Raj}}(2024)}]{2024PhR..1052....1B}%
  \BibitemOpen
  \bibfield  {author} {\bibinfo {author} {\bibfnamefont {J.}~\bibnamefont
  {{Bramante}}}\ and\ \bibinfo {author} {\bibfnamefont {N.}~\bibnamefont
  {{Raj}}},\ }\bibfield  {title} {\bibinfo {title} {{Dark matter in compact
  stars}},\ }\href {https://doi.org/10.1016/j.physrep.2023.12.001} {\bibfield
  {journal} {\bibinfo  {journal} {\physrep}\ }\textbf {\bibinfo {volume}
  {1052}},\ \bibinfo {pages} {1} (\bibinfo {year} {2024})}\BibitemShut
  {NoStop}%
\bibitem [{\citenamefont {{Cardoso}}\ and\ \citenamefont
  {{Pani}}(2019)}]{2019LRR....22....4C}%
  \BibitemOpen
  \bibfield  {author} {\bibinfo {author} {\bibfnamefont {V.}~\bibnamefont
  {{Cardoso}}}\ and\ \bibinfo {author} {\bibfnamefont {P.}~\bibnamefont
  {{Pani}}},\ }\bibfield  {title} {\bibinfo {title} {{Testing the nature of
  dark compact objects: a status report}},\ }\href
  {https://doi.org/10.1007/s41114-019-0020-4} {\bibfield  {journal} {\bibinfo
  {journal} {\lrr}\ }\textbf {\bibinfo {volume} {22}},\ \bibinfo {pages} {4}
  (\bibinfo {year} {2019})}\BibitemShut {NoStop}%
\bibitem [{\citenamefont {{Tulin}}\ and\ \citenamefont
  {{Yu}}(2018)}]{2018PhR...730....1T}%
  \BibitemOpen
  \bibfield  {author} {\bibinfo {author} {\bibfnamefont {S.}~\bibnamefont
  {{Tulin}}}\ and\ \bibinfo {author} {\bibfnamefont {H.-B.}\ \bibnamefont
  {{Yu}}},\ }\bibfield  {title} {\bibinfo {title} {{Dark matter
  self-interactions and small scale structure}},\ }\href
  {https://doi.org/10.1016/j.physrep.2017.11.004} {\bibfield  {journal}
  {\bibinfo  {journal} {\physrep}\ }\textbf {\bibinfo {volume} {730}},\
  \bibinfo {pages} {1} (\bibinfo {year} {2018})}\BibitemShut {NoStop}%
\bibitem [{\citenamefont {{Wise}}\ and\ \citenamefont
  {{Zhang}}(2014)}]{2014PhRvD..90e5030W}%
  \BibitemOpen
  \bibfield  {author} {\bibinfo {author} {\bibfnamefont {M.~B.}\ \bibnamefont
  {{Wise}}}\ and\ \bibinfo {author} {\bibfnamefont {Y.}~\bibnamefont
  {{Zhang}}},\ }\bibfield  {title} {\bibinfo {title} {{Stable bound states of
  asymmetric dark matter}},\ }\href
  {https://doi.org/10.1103/PhysRevD.90.055030} {\bibfield  {journal} {\bibinfo
  {journal} {\prd}\ }\textbf {\bibinfo {volume} {90}},\ \bibinfo {pages}
  {055030} (\bibinfo {year} {2014})}\BibitemShut {NoStop}%
\bibitem [{\citenamefont {{Gresham}}\ \emph {et~al.}(2017)\citenamefont
  {{Gresham}}, \citenamefont {{Lou}},\ and\ \citenamefont
  {{Zurek}}}]{2017PhRvD..96i6012G}%
  \BibitemOpen
  \bibfield  {author} {\bibinfo {author} {\bibfnamefont {M.~I.}\ \bibnamefont
  {{Gresham}}}, \bibinfo {author} {\bibfnamefont {H.~K.}\ \bibnamefont
  {{Lou}}},\ and\ \bibinfo {author} {\bibfnamefont {K.~M.}\ \bibnamefont
  {{Zurek}}},\ }\bibfield  {title} {\bibinfo {title} {{Nuclear structure of
  bound states of asymmetric dark matter}},\ }\href
  {https://doi.org/10.1103/PhysRevD.96.096012} {\bibfield  {journal} {\bibinfo
  {journal} {\prd}\ }\textbf {\bibinfo {volume} {96}},\ \bibinfo {pages}
  {096012} (\bibinfo {year} {2017})}\BibitemShut {NoStop}%
\bibitem [{\citenamefont {{Narain}}\ \emph {et~al.}(2006)\citenamefont
  {{Narain}}, \citenamefont {{Schaffner-Bielich}},\ and\ \citenamefont
  {{Mishustin}}}]{2006PhRvD..74f3003N}%
  \BibitemOpen
  \bibfield  {author} {\bibinfo {author} {\bibfnamefont {G.}~\bibnamefont
  {{Narain}}}, \bibinfo {author} {\bibfnamefont {J.}~\bibnamefont
  {{Schaffner-Bielich}}},\ and\ \bibinfo {author} {\bibfnamefont {I.~N.}\
  \bibnamefont {{Mishustin}}},\ }\bibfield  {title} {\bibinfo {title} {{Compact
  stars made of fermionic dark matter}},\ }\href
  {https://doi.org/10.1103/PhysRevD.74.063003} {\bibfield  {journal} {\bibinfo
  {journal} {\prd}\ }\textbf {\bibinfo {volume} {74}},\ \bibinfo {pages}
  {063003} (\bibinfo {year} {2006})}\BibitemShut {NoStop}%
\bibitem [{\citenamefont {{Kouvaris}}\ and\ \citenamefont
  {{Nielsen}}(2015)}]{2015PhRvD..92f3526K}%
  \BibitemOpen
  \bibfield  {author} {\bibinfo {author} {\bibfnamefont {C.}~\bibnamefont
  {{Kouvaris}}}\ and\ \bibinfo {author} {\bibfnamefont {N.~G.}\ \bibnamefont
  {{Nielsen}}},\ }\bibfield  {title} {\bibinfo {title} {{Asymmetric dark matter
  stars}},\ }\href {https://doi.org/10.1103/PhysRevD.92.063526} {\bibfield
  {journal} {\bibinfo  {journal} {\prd}\ }\textbf {\bibinfo {volume} {92}},\
  \bibinfo {pages} {063526} (\bibinfo {year} {2015})}\BibitemShut {NoStop}%
\bibitem [{\citenamefont {{Abbott}}\ \emph {et~al.}(2017)\citenamefont
  {{Abbott}} \emph {et~al.}}]{2017PhRvL.119p1101A}%
  \BibitemOpen
  \bibfield  {author} {\bibinfo {author} {\bibfnamefont {B.~P.}\ \bibnamefont
  {{Abbott}}} \emph {et~al.},\ }\bibfield  {title} {\bibinfo {title}
  {{GW170817: Observation of gravitational waves from a binary neutron star
  inspiral}},\ }\href {https://doi.org/10.1103/PhysRevLett.119.161101}
  {\bibfield  {journal} {\bibinfo  {journal} {\prl}\ }\textbf {\bibinfo
  {volume} {119}},\ \bibinfo {pages} {161101} (\bibinfo {year}
  {2017})}\BibitemShut {NoStop}%
\bibitem [{\citenamefont {{Abbott}}\ \emph {et~al.}(2018)\citenamefont
  {{Abbott}}, \citenamefont {{Abbott}}, \citenamefont {{Abbott}} \emph
  {et~al.}}]{2018PhRvL.121p1101A}%
  \BibitemOpen
  \bibfield  {author} {\bibinfo {author} {\bibfnamefont {B.~P.}\ \bibnamefont
  {{Abbott}}}, \bibinfo {author} {\bibfnamefont {R.}~\bibnamefont {{Abbott}}},
  \bibinfo {author} {\bibfnamefont {T.~D.}\ \bibnamefont {{Abbott}}}, \emph
  {et~al.},\ }\bibfield  {title} {\bibinfo {title} {{GW170817: Measurements of
  Neutron Star Radii and Equation of State}},\ }\href
  {https://doi.org/10.1103/PhysRevLett.121.161101} {\bibfield  {journal}
  {\bibinfo  {journal} {Phys.\ Rev.\ Lett.}\ }\textbf {\bibinfo {volume}
  {121}},\ \bibinfo {pages} {161101} (\bibinfo {year} {2018})}\BibitemShut
  {NoStop}%
\bibitem [{\citenamefont {{Miller}}\ \emph {et~al.}(2021)\citenamefont
  {{Miller}}, \citenamefont {{Lamb}}, \citenamefont {{Dittmann}}, \citenamefont
  {{Bogdanov}}, \citenamefont {{Arzoumanian}}, \citenamefont {{Gendreau}},
  \citenamefont {{Guillot}}, \citenamefont {{Ho}}, \citenamefont {{Lattimer}},
  \citenamefont {{Loewenstein}}, \citenamefont {{Morsink}}, \citenamefont
  {{Ray}}, \citenamefont {{Wolff}}, \citenamefont {{Baker}}, \citenamefont
  {{Cazeau}}, \citenamefont {{Manthripragada}}, \citenamefont {{Markwardt}},
  \citenamefont {{Okajima}}, \citenamefont {{Pollard}}, \citenamefont
  {{Stairs}}, \citenamefont {{Stroh}},\ and\ \citenamefont
  {{Tamba}}}]{2021ApJ...918L..28M}%
  \BibitemOpen
  \bibfield  {author} {\bibinfo {author} {\bibfnamefont {M.~C.}\ \bibnamefont
  {{Miller}}}, \bibinfo {author} {\bibfnamefont {F.~K.}\ \bibnamefont
  {{Lamb}}}, \bibinfo {author} {\bibfnamefont {A.~J.}\ \bibnamefont
  {{Dittmann}}}, \bibinfo {author} {\bibfnamefont {S.}~\bibnamefont
  {{Bogdanov}}}, \bibinfo {author} {\bibfnamefont {Z.}~\bibnamefont
  {{Arzoumanian}}}, \bibinfo {author} {\bibfnamefont {K.~C.}\ \bibnamefont
  {{Gendreau}}}, \bibinfo {author} {\bibfnamefont {S.}~\bibnamefont
  {{Guillot}}}, \bibinfo {author} {\bibfnamefont {W.~C.~G.}\ \bibnamefont
  {{Ho}}}, \bibinfo {author} {\bibfnamefont {J.~M.}\ \bibnamefont
  {{Lattimer}}}, \bibinfo {author} {\bibfnamefont {M.}~\bibnamefont
  {{Loewenstein}}}, \bibinfo {author} {\bibfnamefont {S.~M.}\ \bibnamefont
  {{Morsink}}}, \bibinfo {author} {\bibfnamefont {P.~S.}\ \bibnamefont
  {{Ray}}}, \bibinfo {author} {\bibfnamefont {M.~T.}\ \bibnamefont {{Wolff}}},
  \bibinfo {author} {\bibfnamefont {C.~L.}\ \bibnamefont {{Baker}}}, \bibinfo
  {author} {\bibfnamefont {T.}~\bibnamefont {{Cazeau}}}, \bibinfo {author}
  {\bibfnamefont {S.}~\bibnamefont {{Manthripragada}}}, \bibinfo {author}
  {\bibfnamefont {C.~B.}\ \bibnamefont {{Markwardt}}}, \bibinfo {author}
  {\bibfnamefont {T.}~\bibnamefont {{Okajima}}}, \bibinfo {author}
  {\bibfnamefont {S.}~\bibnamefont {{Pollard}}}, \bibinfo {author}
  {\bibfnamefont {I.}~\bibnamefont {{Stairs}}}, \bibinfo {author}
  {\bibfnamefont {M.~C.}\ \bibnamefont {{Stroh}}},\ and\ \bibinfo {author}
  {\bibfnamefont {T.}~\bibnamefont {{Tamba}}},\ }\bibfield  {title} {\bibinfo
  {title} {{The Radius of PSR J0740+6620 from NICER and XMM-Newton Data}},\
  }\href {https://doi.org/10.3847/2041-8213/ac089b} {\bibfield  {journal}
  {\bibinfo  {journal} {\apjl}\ }\textbf {\bibinfo {volume} {918}},\ \bibinfo
  {pages} {L28} (\bibinfo {year} {2021})}\BibitemShut {NoStop}%
\bibitem [{\citenamefont {{Lopes}}(2026)}]{LopesQuantumStarsNewtonian2026}%
  \BibitemOpen
  \bibfield  {author} {\bibinfo {author} {\bibfnamefont {I.}~\bibnamefont
  {{Lopes}}},\ }\bibfield  {title} {\bibinfo {title} {{Self-gravitating quantum
  stars with a globally relevant Bohm potential}},\ }\href
  {https://doi.org/10.1103/m6jf-8cyl} {\bibfield  {journal} {\bibinfo
  {journal} {\prd}\ }\textbf {\bibinfo {volume} {113}},\ \bibinfo {pages}
  {115043} (\bibinfo {year} {2026})},\ \Eprint
  {https://arxiv.org/abs/2606.01103} {arXiv:2606.01103} \BibitemShut {NoStop}%
\bibitem [{\citenamefont {{Walecka}}(1974)}]{1974AnPhy..83..491W}%
  \BibitemOpen
  \bibfield  {author} {\bibinfo {author} {\bibfnamefont {J.~D.}\ \bibnamefont
  {{Walecka}}},\ }\bibfield  {title} {\bibinfo {title} {{A theory of highly
  condensed matter}},\ }\href {https://doi.org/10.1016/0003-4916(74)90208-5}
  {\bibfield  {journal} {\bibinfo  {journal} {Ann.~Phys.}\ }\textbf {\bibinfo
  {volume} {83}},\ \bibinfo {pages} {491} (\bibinfo {year} {1974})}\BibitemShut
  {NoStop}%
\bibitem [{\citenamefont {{Serot}}\ and\ \citenamefont
  {{Walecka}}(1997)}]{1997IJMPE...6..515S}%
  \BibitemOpen
  \bibfield  {author} {\bibinfo {author} {\bibfnamefont {B.~D.}\ \bibnamefont
  {{Serot}}}\ and\ \bibinfo {author} {\bibfnamefont {J.~D.}\ \bibnamefont
  {{Walecka}}},\ }\bibfield  {title} {\bibinfo {title} {{Recent Progress in
  Quantum Hadrodynamics}},\ }\href {https://doi.org/10.1142/S0218301397000299}
  {\bibfield  {journal} {\bibinfo  {journal} {\ijmpe}\ }\textbf {\bibinfo
  {volume} {6}},\ \bibinfo {pages} {515} (\bibinfo {year} {1997})}\BibitemShut
  {NoStop}%
\bibitem [{\citenamefont {{Boguta}}\ and\ \citenamefont
  {{Bodmer}}(1977)}]{1977NuPhA.292..413B}%
  \BibitemOpen
  \bibfield  {author} {\bibinfo {author} {\bibfnamefont {J.}~\bibnamefont
  {{Boguta}}}\ and\ \bibinfo {author} {\bibfnamefont {A.~R.}\ \bibnamefont
  {{Bodmer}}},\ }\bibfield  {title} {\bibinfo {title} {{Relativistic
  calculation of nuclear matter and the nuclear surface}},\ }\href
  {https://doi.org/10.1016/0375-9474(77)90626-1} {\bibfield  {journal}
  {\bibinfo  {journal} {Nucl.~Phys.~A}\ }\textbf {\bibinfo {volume} {292}},\
  \bibinfo {pages} {413} (\bibinfo {year} {1977})}\BibitemShut {NoStop}%
\bibitem [{\citenamefont {{Oppenheimer}}\ and\ \citenamefont
  {{Volkoff}}(1939)}]{1939PhRv...55..374O}%
  \BibitemOpen
  \bibfield  {author} {\bibinfo {author} {\bibfnamefont {J.~R.}\ \bibnamefont
  {{Oppenheimer}}}\ and\ \bibinfo {author} {\bibfnamefont {G.~M.}\ \bibnamefont
  {{Volkoff}}},\ }\bibfield  {title} {\bibinfo {title} {{On Massive Neutron
  Cores}},\ }\href {https://doi.org/10.1103/PhysRev.55.374} {\bibfield
  {journal} {\bibinfo  {journal} {Phys.~Rev.}\ }\textbf {\bibinfo {volume}
  {55}},\ \bibinfo {pages} {374} (\bibinfo {year} {1939})}\BibitemShut
  {NoStop}%
\bibitem [{\citenamefont {{Misner}}\ \emph {et~al.}(1973)\citenamefont
  {{Misner}}, \citenamefont {{Thorne}},\ and\ \citenamefont
  {{Wheeler}}}]{1973grav.book.....M}%
  \BibitemOpen
  \bibfield  {author} {\bibinfo {author} {\bibfnamefont {C.~W.}\ \bibnamefont
  {{Misner}}}, \bibinfo {author} {\bibfnamefont {K.~S.}\ \bibnamefont
  {{Thorne}}},\ and\ \bibinfo {author} {\bibfnamefont {J.~A.}\ \bibnamefont
  {{Wheeler}}},\ }\href
  {https://ui.adsabs.harvard.edu/abs/1973grav.book.....M/abstract} {\emph
  {\bibinfo {title} {{Gravitation}}}}\ (\bibinfo  {publisher} {W.~H.~Freeman},\
  \bibinfo {address} {San Francisco},\ \bibinfo {year} {1973})\BibitemShut
  {NoStop}%
\bibitem [{\citenamefont {{Birrell}}\ and\ \citenamefont
  {{Davies}}(1982)}]{1996qtf..book.....B}%
  \BibitemOpen
  \bibfield  {author} {\bibinfo {author} {\bibfnamefont {N.~D.}\ \bibnamefont
  {{Birrell}}}\ and\ \bibinfo {author} {\bibfnamefont {P.~C.~W.}\ \bibnamefont
  {{Davies}}},\ }\href {https://doi.org/10.1017/CBO9780511622632} {\emph
  {\bibinfo {title} {{Quantum Fields in Curved Space}}}}\ (\bibinfo
  {publisher} {Cambridge University Press},\ \bibinfo {year}
  {1982})\BibitemShut {NoStop}%
\bibitem [{\citenamefont {{Parker}}\ and\ \citenamefont
  {{Toms}}(2009)}]{2009qftc.book.....P}%
  \BibitemOpen
  \bibfield  {author} {\bibinfo {author} {\bibfnamefont {L.~E.}\ \bibnamefont
  {{Parker}}}\ and\ \bibinfo {author} {\bibfnamefont {D.~J.}\ \bibnamefont
  {{Toms}}},\ }\href {https://doi.org/10.1017/CBO9780511813924} {\emph
  {\bibinfo {title} {{Quantum Field Theory in Curved Spacetime: Quantized
  Fields and Gravity}}}}\ (\bibinfo  {publisher} {Cambridge University Press},\
  \bibinfo {year} {2009})\BibitemShut {NoStop}%
\bibitem [{\citenamefont {{Glendenning}}(1997)}]{1997csnp.book.....G}%
  \BibitemOpen
  \bibfield  {author} {\bibinfo {author} {\bibfnamefont {N.~K.}\ \bibnamefont
  {{Glendenning}}},\ }\href {https://doi.org/10.1007/978-1-4684-0491-3} {\emph
  {\bibinfo {title} {{Compact Stars: Nuclear Physics, Particle Physics, and
  General Relativity}}}}\ (\bibinfo  {publisher} {Springer, New York},\
  \bibinfo {year} {1997})\BibitemShut {NoStop}%
\bibitem [{\citenamefont {{Proca}}(1936)}]{Proca1936}%
  \BibitemOpen
  \bibfield  {author} {\bibinfo {author} {\bibfnamefont {A.}~\bibnamefont
  {{Proca}}},\ }\bibfield  {title} {\bibinfo {title} {{Sur la th\'eorie
  ondulatoire des \'electrons positifs et n\'egatifs}},\ }\href
  {https://doi.org/10.1051/jphysrad:0193600708034700} {\bibfield  {journal}
  {\bibinfo  {journal} {J.~Phys.~Radium}\ }\textbf {\bibinfo {volume} {7}},\
  \bibinfo {pages} {347} (\bibinfo {year} {1936})}\BibitemShut {NoStop}%
\bibitem [{\citenamefont {{Shapiro}}\ and\ \citenamefont
  {{Teukolsky}}(1983)}]{1983bhwd.book.....S}%
  \BibitemOpen
  \bibfield  {author} {\bibinfo {author} {\bibfnamefont {S.~L.}\ \bibnamefont
  {{Shapiro}}}\ and\ \bibinfo {author} {\bibfnamefont {S.~A.}\ \bibnamefont
  {{Teukolsky}}},\ }\href {https://doi.org/10.1002/9783527617661} {\emph
  {\bibinfo {title} {{Black Holes, White Dwarfs, and Neutron Stars: The Physics
  of Compact Objects}}}}\ (\bibinfo  {publisher} {Wiley},\ \bibinfo {address}
  {New York},\ \bibinfo {year} {1983})\BibitemShut {NoStop}%
\bibitem [{\citenamefont {{von Weizs{\"a}cker}}(1935)}]{1935ZPhy...96..431W}%
  \BibitemOpen
  \bibfield  {author} {\bibinfo {author} {\bibfnamefont {C.~F.}\ \bibnamefont
  {{von Weizs{\"a}cker}}},\ }\bibfield  {title} {\bibinfo {title} {{Zur Theorie
  der Kernmassen}},\ }\href {https://doi.org/10.1007/BF01337700} {\bibfield
  {journal} {\bibinfo  {journal} {\zphy}\ }\textbf {\bibinfo {volume} {96}},\
  \bibinfo {pages} {431} (\bibinfo {year} {1935})}\BibitemShut {NoStop}%
\bibitem [{\citenamefont {{Kirzhnits}}(1957)}]{1957JETP....5...64K}%
  \BibitemOpen
  \bibfield  {author} {\bibinfo {author} {\bibfnamefont {D.~A.}\ \bibnamefont
  {{Kirzhnits}}},\ }\bibfield  {title} {\bibinfo {title} {{Quantum corrections
  to the Thomas--Fermi equation}},\ }\href
  {http://jetp.ras.ru/cgi-bin/e/index/e/5/1/p64?a=list} {\bibfield  {journal}
  {\bibinfo  {journal} {Sov.~Phys.~JETP}\ }\textbf {\bibinfo {volume} {5}},\
  \bibinfo {pages} {64} (\bibinfo {year} {1957})}\BibitemShut {NoStop}%
\bibitem [{\citenamefont {{Moldabekov}}\ \emph {et~al.}(2018)\citenamefont
  {{Moldabekov}}, \citenamefont {{Bonitz}},\ and\ \citenamefont
  {{Ramazanov}}}]{2018PhPl...25c1903M}%
  \BibitemOpen
  \bibfield  {author} {\bibinfo {author} {\bibfnamefont {Z.~A.}\ \bibnamefont
  {{Moldabekov}}}, \bibinfo {author} {\bibfnamefont {M.}~\bibnamefont
  {{Bonitz}}},\ and\ \bibinfo {author} {\bibfnamefont {T.~S.}\ \bibnamefont
  {{Ramazanov}}},\ }\bibfield  {title} {\bibinfo {title} {{Theoretical
  foundations of quantum hydrodynamics for plasmas}},\ }\href
  {https://doi.org/10.1063/1.5003910} {\bibfield  {journal} {\bibinfo
  {journal} {\pop}\ }\textbf {\bibinfo {volume} {25}},\ \bibinfo {pages}
  {031903} (\bibinfo {year} {2018})}\BibitemShut {NoStop}%
\bibitem [{\citenamefont {{Tolman}}(1930)}]{1930PhRv...35..904T}%
  \BibitemOpen
  \bibfield  {author} {\bibinfo {author} {\bibfnamefont {R.~C.}\ \bibnamefont
  {{Tolman}}},\ }\bibfield  {title} {\bibinfo {title} {{On the Weight of Heat
  and Thermal Equilibrium in General Relativity}},\ }\href
  {https://doi.org/10.1103/PhysRev.35.904} {\bibfield  {journal} {\bibinfo
  {journal} {Phys.~Rev.}\ }\textbf {\bibinfo {volume} {35}},\ \bibinfo {pages}
  {904} (\bibinfo {year} {1930})}\BibitemShut {NoStop}%
\bibitem [{\citenamefont {{Klein}}(1949)}]{1949RvMP...21..531K}%
  \BibitemOpen
  \bibfield  {author} {\bibinfo {author} {\bibfnamefont {O.}~\bibnamefont
  {{Klein}}},\ }\bibfield  {title} {\bibinfo {title} {{On the Thermodynamical
  Equilibrium of Fluids in Gravitational Fields}},\ }\href
  {https://doi.org/10.1103/RevModPhys.21.531} {\bibfield  {journal} {\bibinfo
  {journal} {Rev.~Mod.~Phys.}\ }\textbf {\bibinfo {volume} {21}},\ \bibinfo
  {pages} {531} (\bibinfo {year} {1949})}\BibitemShut {NoStop}%
\bibitem [{\citenamefont {{Buchdahl}}(1959)}]{1959PhRv..116.1027B}%
  \BibitemOpen
  \bibfield  {author} {\bibinfo {author} {\bibfnamefont {H.~A.}\ \bibnamefont
  {{Buchdahl}}},\ }\bibfield  {title} {\bibinfo {title} {{General Relativistic
  Fluid Spheres}},\ }\href {https://doi.org/10.1103/PhysRev.116.1027}
  {\bibfield  {journal} {\bibinfo  {journal} {Phys.\ Rev.}\ }\textbf {\bibinfo
  {volume} {116}},\ \bibinfo {pages} {1027} (\bibinfo {year}
  {1959})}\BibitemShut {NoStop}%
\bibitem [{\citenamefont {{Virtanen}}\ \emph {et~al.}(2020)\citenamefont
  {{Virtanen}}, \citenamefont {{Gommers}}, \citenamefont {{Oliphant}} \emph
  {et~al.}}]{2020NatMe..17..261V}%
  \BibitemOpen
  \bibfield  {author} {\bibinfo {author} {\bibfnamefont {P.}~\bibnamefont
  {{Virtanen}}}, \bibinfo {author} {\bibfnamefont {R.}~\bibnamefont
  {{Gommers}}}, \bibinfo {author} {\bibfnamefont {T.~E.}\ \bibnamefont
  {{Oliphant}}}, \emph {et~al.},\ }\bibfield  {title} {\bibinfo {title} {{SciPy
  1.0: fundamental algorithms for scientific computing in Python}},\ }\href
  {https://doi.org/10.1038/s41592-019-0686-2} {\bibfield  {journal} {\bibinfo
  {journal} {Nature Methods}\ }\textbf {\bibinfo {volume} {17}},\ \bibinfo
  {pages} {261} (\bibinfo {year} {2020})}\BibitemShut {NoStop}%
\bibitem [{\citenamefont {{Hinderer}}(2008)}]{2008ApJ...677.1216H}%
  \BibitemOpen
  \bibfield  {author} {\bibinfo {author} {\bibfnamefont {T.}~\bibnamefont
  {{Hinderer}}},\ }\bibfield  {title} {\bibinfo {title} {{Tidal Love numbers of
  neutron stars}},\ }\href {https://doi.org/10.1086/533487} {\bibfield
  {journal} {\bibinfo  {journal} {\apj}\ }\textbf {\bibinfo {volume} {677}},\
  \bibinfo {pages} {1216} (\bibinfo {year} {2008})}\BibitemShut {NoStop}%
\bibitem [{\citenamefont {{Flanagan}}\ and\ \citenamefont
  {{Hinderer}}(2008)}]{2008PhRvD..77b1502F}%
  \BibitemOpen
  \bibfield  {author} {\bibinfo {author} {\bibfnamefont {{\'E}.~{\'E}.}\
  \bibnamefont {{Flanagan}}}\ and\ \bibinfo {author} {\bibfnamefont
  {T.}~\bibnamefont {{Hinderer}}},\ }\bibfield  {title} {\bibinfo {title}
  {{Constraining neutron-star tidal Love numbers with gravitational-wave
  detectors}},\ }\href {https://doi.org/10.1103/PhysRevD.77.021502} {\bibfield
  {journal} {\bibinfo  {journal} {\prd}\ }\textbf {\bibinfo {volume} {77}},\
  \bibinfo {pages} {021502} (\bibinfo {year} {2008})}\BibitemShut {NoStop}%
\bibitem [{\citenamefont {{Lee}}\ \emph
  {et~al.}(1977{\natexlab{a}})\citenamefont {{Lee}}, \citenamefont {{Quigg}},\
  and\ \citenamefont {{Thacker}}}]{1977PhRvL..38..883L}%
  \BibitemOpen
  \bibfield  {author} {\bibinfo {author} {\bibfnamefont {B.~W.}\ \bibnamefont
  {{Lee}}}, \bibinfo {author} {\bibfnamefont {C.}~\bibnamefont {{Quigg}}},\
  and\ \bibinfo {author} {\bibfnamefont {H.~B.}\ \bibnamefont {{Thacker}}},\
  }\bibfield  {title} {\bibinfo {title} {{Strength of Weak Interactions at Very
  High Energies and the Higgs Boson Mass}},\ }\href
  {https://doi.org/10.1103/PhysRevLett.38.883} {\bibfield  {journal} {\bibinfo
  {journal} {\prl}\ }\textbf {\bibinfo {volume} {38}},\ \bibinfo {pages} {883}
  (\bibinfo {year} {1977}{\natexlab{a}})}\BibitemShut {NoStop}%
\bibitem [{\citenamefont {{Lee}}\ \emph
  {et~al.}(1977{\natexlab{b}})\citenamefont {{Lee}}, \citenamefont {{Quigg}},\
  and\ \citenamefont {{Thacker}}}]{1977PhRvD..16.1519L}%
  \BibitemOpen
  \bibfield  {author} {\bibinfo {author} {\bibfnamefont {B.~W.}\ \bibnamefont
  {{Lee}}}, \bibinfo {author} {\bibfnamefont {C.}~\bibnamefont {{Quigg}}},\
  and\ \bibinfo {author} {\bibfnamefont {H.~B.}\ \bibnamefont {{Thacker}}},\
  }\bibfield  {title} {\bibinfo {title} {{Weak Interactions at Very High
  Energies: The Role of the Higgs-Boson Mass}},\ }\href
  {https://doi.org/10.1103/PhysRevD.16.1519} {\bibfield  {journal} {\bibinfo
  {journal} {\prd}\ }\textbf {\bibinfo {volume} {16}},\ \bibinfo {pages} {1519}
  (\bibinfo {year} {1977}{\natexlab{b}})}\BibitemShut {NoStop}%
\bibitem [{\citenamefont {{Goodsell}}\ and\ \citenamefont
  {{Staub}}(2018)}]{2018EPJC...78..748G}%
  \BibitemOpen
  \bibfield  {author} {\bibinfo {author} {\bibfnamefont {M.~D.}\ \bibnamefont
  {{Goodsell}}}\ and\ \bibinfo {author} {\bibfnamefont {F.}~\bibnamefont
  {{Staub}}},\ }\bibfield  {title} {\bibinfo {title} {{Unitarity constraints on
  general scalar couplings with SARAH}},\ }\href
  {https://doi.org/10.1140/epjc/s10052-018-6127-z} {\bibfield  {journal}
  {\bibinfo  {journal} {Eur.~Phys.~J.~C}\ }\textbf {\bibinfo {volume} {78}},\
  \bibinfo {pages} {748} (\bibinfo {year} {2018})},\ \Eprint
  {https://arxiv.org/abs/1805.07306} {arXiv:1805.07306} \BibitemShut {NoStop}%
\bibitem [{\citenamefont {{Witten}}(1984)}]{1984PhRvD..30..272W}%
  \BibitemOpen
  \bibfield  {author} {\bibinfo {author} {\bibfnamefont {E.}~\bibnamefont
  {{Witten}}},\ }\bibfield  {title} {\bibinfo {title} {{Cosmic separation of
  phases}},\ }\href {https://doi.org/10.1103/PhysRevD.30.272} {\bibfield
  {journal} {\bibinfo  {journal} {\prd}\ }\textbf {\bibinfo {volume} {30}},\
  \bibinfo {pages} {272} (\bibinfo {year} {1984})}\BibitemShut {NoStop}%
\bibitem [{\citenamefont {{Alcock}}\ \emph {et~al.}(1986)\citenamefont
  {{Alcock}}, \citenamefont {{Farhi}},\ and\ \citenamefont
  {{Olinto}}}]{1986ApJ...310..261A}%
  \BibitemOpen
  \bibfield  {author} {\bibinfo {author} {\bibfnamefont {C.}~\bibnamefont
  {{Alcock}}}, \bibinfo {author} {\bibfnamefont {E.}~\bibnamefont {{Farhi}}},\
  and\ \bibinfo {author} {\bibfnamefont {A.}~\bibnamefont {{Olinto}}},\
  }\bibfield  {title} {\bibinfo {title} {{Strange stars}},\ }\href
  {https://doi.org/10.1086/164679} {\bibfield  {journal} {\bibinfo  {journal}
  {\apj}\ }\textbf {\bibinfo {volume} {310}},\ \bibinfo {pages} {261} (\bibinfo
  {year} {1986})}\BibitemShut {NoStop}%
\bibitem [{\citenamefont {{Haensel}}\ \emph {et~al.}(1986)\citenamefont
  {{Haensel}}, \citenamefont {{Zdunik}},\ and\ \citenamefont
  {{Schaeffer}}}]{1986A&A...160..121H}%
  \BibitemOpen
  \bibfield  {author} {\bibinfo {author} {\bibfnamefont {P.}~\bibnamefont
  {{Haensel}}}, \bibinfo {author} {\bibfnamefont {J.~L.}\ \bibnamefont
  {{Zdunik}}},\ and\ \bibinfo {author} {\bibfnamefont {R.}~\bibnamefont
  {{Schaeffer}}},\ }\bibfield  {title} {\bibinfo {title} {{Strange quark
  stars}},\ }\href@noop {} {\bibfield  {journal} {\bibinfo  {journal} {\aap}\
  }\textbf {\bibinfo {volume} {160}},\ \bibinfo {pages} {121} (\bibinfo {year}
  {1986})}\BibitemShut {NoStop}%
\bibitem [{\citenamefont {{Wise}}\ and\ \citenamefont
  {{Zhang}}(2015)}]{2015JHEP...02..023W}%
  \BibitemOpen
  \bibfield  {author} {\bibinfo {author} {\bibfnamefont {M.~B.}\ \bibnamefont
  {{Wise}}}\ and\ \bibinfo {author} {\bibfnamefont {Y.}~\bibnamefont
  {{Zhang}}},\ }\bibfield  {title} {\bibinfo {title} {{Yukawa bound states of a
  large number of fermions}},\ }\href {https://doi.org/10.1007/JHEP02(2015)023}
  {\bibfield  {journal} {\bibinfo  {journal} {\jhep}\ }\textbf {\bibinfo
  {volume} {02}},\ \bibinfo {pages} {023} (\bibinfo {year} {2015})}\BibitemShut
  {NoStop}%
\bibitem [{\citenamefont {{Gresham}}\ \emph {et~al.}(2018)\citenamefont
  {{Gresham}}, \citenamefont {{Lou}},\ and\ \citenamefont
  {{Zurek}}}]{2018PhRvD..98i6001G}%
  \BibitemOpen
  \bibfield  {author} {\bibinfo {author} {\bibfnamefont {M.~I.}\ \bibnamefont
  {{Gresham}}}, \bibinfo {author} {\bibfnamefont {H.~K.}\ \bibnamefont
  {{Lou}}},\ and\ \bibinfo {author} {\bibfnamefont {K.~M.}\ \bibnamefont
  {{Zurek}}},\ }\bibfield  {title} {\bibinfo {title} {{Astrophysical signatures
  of asymmetric dark matter bound states}},\ }\href
  {https://doi.org/10.1103/PhysRevD.98.096001} {\bibfield  {journal} {\bibinfo
  {journal} {\prd}\ }\textbf {\bibinfo {volume} {98}},\ \bibinfo {pages}
  {096001} (\bibinfo {year} {2018})}\BibitemShut {NoStop}%
\bibitem [{\citenamefont {{Hardy}}\ \emph {et~al.}(2015)\citenamefont
  {{Hardy}}, \citenamefont {{Lasenby}}, \citenamefont {{March-Russell}},\ and\
  \citenamefont {{West}}}]{2015JHEP...06..011H}%
  \BibitemOpen
  \bibfield  {author} {\bibinfo {author} {\bibfnamefont {E.}~\bibnamefont
  {{Hardy}}}, \bibinfo {author} {\bibfnamefont {R.}~\bibnamefont {{Lasenby}}},
  \bibinfo {author} {\bibfnamefont {J.}~\bibnamefont {{March-Russell}}},\ and\
  \bibinfo {author} {\bibfnamefont {S.~M.}\ \bibnamefont {{West}}},\ }\bibfield
   {title} {\bibinfo {title} {{Big bang synthesis of nuclear dark matter}},\
  }\href {https://doi.org/10.1007/JHEP06(2015)011} {\bibfield  {journal}
  {\bibinfo  {journal} {\jhep}\ }\textbf {\bibinfo {volume} {06}},\ \bibinfo
  {pages} {011} (\bibinfo {year} {2015})}\BibitemShut {NoStop}%
\bibitem [{\citenamefont {{Panotopoulos}}\ and\ \citenamefont
  {{Lopes}}(2017)}]{2017PhRvD..96h3004P}%
  \BibitemOpen
  \bibfield  {author} {\bibinfo {author} {\bibfnamefont {G.}~\bibnamefont
  {{Panotopoulos}}}\ and\ \bibinfo {author} {\bibfnamefont {I.}~\bibnamefont
  {{Lopes}}},\ }\bibfield  {title} {\bibinfo {title} {{Dark matter effect on
  realistic equation of state in neutron stars}},\ }\href
  {https://doi.org/10.1103/PhysRevD.96.083004} {\bibfield  {journal} {\bibinfo
  {journal} {\prd}\ }\textbf {\bibinfo {volume} {96}},\ \bibinfo {pages}
  {083004} (\bibinfo {year} {2017})}\BibitemShut {NoStop}%
\bibitem [{\citenamefont {{Ivanytskyi}}\ \emph {et~al.}(2020)\citenamefont
  {{Ivanytskyi}}, \citenamefont {{Sagun}},\ and\ \citenamefont
  {{Lopes}}}]{2020PhRvD.102f3028I}%
  \BibitemOpen
  \bibfield  {author} {\bibinfo {author} {\bibfnamefont {O.}~\bibnamefont
  {{Ivanytskyi}}}, \bibinfo {author} {\bibfnamefont {V.}~\bibnamefont
  {{Sagun}}},\ and\ \bibinfo {author} {\bibfnamefont {I.}~\bibnamefont
  {{Lopes}}},\ }\bibfield  {title} {\bibinfo {title} {{Neutron stars: New
  constraints on asymmetric dark matter}},\ }\href
  {https://doi.org/10.1103/PhysRevD.102.063028} {\bibfield  {journal} {\bibinfo
   {journal} {\prd}\ }\textbf {\bibinfo {volume} {102}},\ \bibinfo {pages}
  {063028} (\bibinfo {year} {2020})}\BibitemShut {NoStop}%
\bibitem [{\citenamefont {{Punturo}}\ \emph {et~al.}(2010)\citenamefont
  {{Punturo}} \emph {et~al.}}]{2010CQGra..27s4002P}%
  \BibitemOpen
  \bibfield  {author} {\bibinfo {author} {\bibfnamefont {M.}~\bibnamefont
  {{Punturo}}} \emph {et~al.},\ }\bibfield  {title} {\bibinfo {title} {{The
  Einstein Telescope: a third-generation gravitational wave observatory}},\
  }\href {https://doi.org/10.1088/0264-9381/27/19/194002} {\bibfield  {journal}
  {\bibinfo  {journal} {\cqg}\ }\textbf {\bibinfo {volume} {27}},\ \bibinfo
  {pages} {194002} (\bibinfo {year} {2010})}\BibitemShut {NoStop}%
\bibitem [{\citenamefont {{Reitze}}\ \emph {et~al.}(2019)\citenamefont
  {{Reitze}} \emph {et~al.}}]{2019BAAS...51g..35R}%
  \BibitemOpen
  \bibfield  {author} {\bibinfo {author} {\bibfnamefont {D.}~\bibnamefont
  {{Reitze}}} \emph {et~al.},\ }\bibfield  {title} {\bibinfo {title} {{Cosmic
  Explorer: The U.S. Contribution to Gravitational-Wave Astronomy beyond
  LIGO}},\ }\href {https://doi.org/10.48550/arXiv.1907.04833} {\bibfield
  {journal} {\bibinfo  {journal} {Bulletin of the American Astronomical
  Society}\ }\textbf {\bibinfo {volume} {51}},\ \bibinfo {pages} {35} (\bibinfo
  {year} {2019})},\ \Eprint {https://arxiv.org/abs/1907.04833}
  {arXiv:1907.04833} \BibitemShut {NoStop}%
\bibitem [{\citenamefont {{Abbott}}\ \emph {et~al.}(2020)\citenamefont
  {{Abbott}}, \citenamefont {{Abbott}}, \citenamefont {{Abraham}} \emph
  {et~al.}}]{2020ApJ...896L..44A}%
  \BibitemOpen
  \bibfield  {author} {\bibinfo {author} {\bibfnamefont {R.}~\bibnamefont
  {{Abbott}}}, \bibinfo {author} {\bibfnamefont {T.~D.}\ \bibnamefont
  {{Abbott}}}, \bibinfo {author} {\bibfnamefont {S.}~\bibnamefont {{Abraham}}},
  \emph {et~al.},\ }\bibfield  {title} {\bibinfo {title} {{GW190814:
  Gravitational Waves from the Coalescence of a 23 Solar Mass Black Hole with a
  2.6 Solar Mass Compact Object}},\ }\href
  {https://doi.org/10.3847/2041-8213/ab960f} {\bibfield  {journal} {\bibinfo
  {journal} {\apjl}\ }\textbf {\bibinfo {volume} {896}},\ \bibinfo {pages}
  {L44} (\bibinfo {year} {2020})}\BibitemShut {NoStop}%
\bibitem [{\citenamefont {{Tolman}}(1939)}]{1939PhRv...55..364T}%
  \BibitemOpen
  \bibfield  {author} {\bibinfo {author} {\bibfnamefont {R.~C.}\ \bibnamefont
  {{Tolman}}},\ }\bibfield  {title} {\bibinfo {title} {{Static Solutions of
  Einstein's Field Equations for Spheres of Fluid}},\ }\href
  {https://doi.org/10.1103/PhysRev.55.364} {\bibfield  {journal} {\bibinfo
  {journal} {Phys.~Rev.}\ }\textbf {\bibinfo {volume} {55}},\ \bibinfo {pages}
  {364} (\bibinfo {year} {1939})}\BibitemShut {NoStop}%
\end{thebibliography}%

	\appendix

	\section{Derivation of the species-wise force balance}
	\label{app:tov}

	This appendix derives the quantum-corrected force balance~\eqref{eq:TOV_bohm} from the field equations of Section~\ref{sec:model}, so that the structure problem of this paper is entirely self-contained.
	The logic descends from the Einstein equations,
	\begin{equation}
		G_{\mu\nu} = 8\pi G\,T_{\mu\nu}\,,
		\label{eq:app_einstein}
	\end{equation}
	with $G_{\mu\nu}$ the Einstein tensor and $T_{\mu\nu}$ the total stress-energy tensor of fermions and mediator fields: the contracted Bianchi identity $\nabla_\mu G^{\mu\nu}=0$ enforces $\nabla_\mu T^{\mu\nu}=0$ \citep{1973grav.book.....M}, whose radial component in the metric~\eqref{eq:metric} is the relativistic hydrostatic balance of \citet{1939PhRv...55..364T} and \citet{1939PhRv...55..374O} that underlies Eq.~\eqref{eq:mass_nu} \citep{1983bhwd.book.....S}.
	What requires proof is the finer, species-wise statement~\eqref{eq:TOV_bohm}, in which this total conservation law is distributed amongst the species, with the mediator fields brokering the exchange of four-momentum between them \citep{1997csnp.book.....G}.

	Consider first the classical part.
	The kinetic stress-energy tensor of species $i$ is that of a perfect fluid built on quasiparticles of effective mass $m_i^*$, with energy density and pressure given by Eqs.~\eqref{eq:eps_kin} and \eqref{eq:P_kin}.
	Because the mediators exchange four-momentum with the fermions, this tensor is not conserved separately; its covariant divergence equals the force density that the fields exert on the species,
	\begin{equation}
		\nabla_\mu T^{\mu\nu}_{\mathrm{kin},i}
		= g_{\sigma i}\,n_{s,i}\,\partial^\nu\sigma
		- g_{\omega i}\,\Omega^{\nu\mu}\,j_{i,\mu}\,,
		\label{eq:app_force}
	\end{equation}
	the scalar term descending from the dependence of $m_i^*$ on $\sigma$, with the scalar density as its source, and the vector term being the Lorentz-type force of the Proca field on the conserved current.
	For the static metric~\eqref{eq:metric}, with $j_i^\mu=n_i u^\mu$ and $u^\mu$ the static four-velocity, the radial component of Eq.~\eqref{eq:app_force} reads
	\begin{equation}
		\begin{split}
			\frac{\mathrm{d}P_{\mathrm{kin},i}}{\mathrm{d}r}
			={}& -\bigl(\varepsilon_{\mathrm{kin},i}+P_{\mathrm{kin},i}\bigr)\frac{\mathrm{d}\nu}{\mathrm{d}r}
			+ g_{\sigma i}\,n_{s,i}\,\frac{\mathrm{d}\sigma}{\mathrm{d}r} \\
			& - g_{\omega i}\,n_i\,e^{-\nu}\,\frac{\mathrm{d}\bigl(e^{\nu}\omega_0\bigr)}{\mathrm{d}r}\,,
		\end{split}
		\label{eq:app_radial}
	\end{equation}
	where $\omega_0$ is the locally measured vector potential of Eq.~\eqref{eq:omega_alg}; the combination $e^{-\nu}\,\mathrm{d}(e^{\nu}\omega_0)/\mathrm{d}r$ is the orthonormal-frame Proca electric field, whence the placement of the redshift factors, which in the local-density regime of Section~\ref{sec:meanfield} carry the entire distinction between coordinate and orthonormal components.

	Two identities of the relativistic kinetic gas, which follow directly from Eqs.~\eqref{eq:eps_kin}--\eqref{eq:ns}, organise the bookkeeping:
	\begin{equation}
		\mathrm{d}P_{\mathrm{kin},i} = n_i\,\mathrm{d}E_{F,i} - n_{s,i}\,\mathrm{d}m_i^*\,,
		\qquad
		\varepsilon_{\mathrm{kin},i}+P_{\mathrm{kin},i} = n_i\,E_{F,i}\,,
		\label{eq:app_thermoid}
	\end{equation}
	with $E_{F,i}=\sqrt{p_{F,i}^2+m_i^{*2}}$ the Fermi energy.
	Since $\mathrm{d}m_i^*=-g_{\sigma i}\,\mathrm{d}\sigma$, the scalar force cancels identically between the two sides of Eq.~\eqref{eq:app_radial}, and expanding the vector term collects what remains into the full chemical potential of Eq.~\eqref{eq:mu_i}:
	\begin{equation}
		\frac{\mathrm{d}\mu_i}{\mathrm{d}r} = -\,\mu_i\,\frac{\mathrm{d}\nu}{\mathrm{d}r}\,.
		\label{eq:app_tolman}
	\end{equation}
	The scalar force is thus absorbed by the effective mass and the vector force by the chemical potential, which is the precise sense in which neither requires an explicit term in the fluid equation.
	It is convenient to define the species-wise quantities thermodynamically rather than by any division of the field energies: $P_i$ is obtained by integrating $\mathrm{d}P_i=n_i\,\mathrm{d}\mu_i$ along the equation of state and $\varepsilon_i\equiv\mu_i n_i-P_i$, whence $\varepsilon_i+P_i=\mu_i n_i$ holds by construction, and the sums $\sum_i P_i$ and $\sum_i\varepsilon_i$ reproduce Eqs.~\eqref{eq:eps_tot} and \eqref{eq:P_tot} by virtue of the total identity $P=\sum_i\mu_i n_i-\varepsilon$ quoted in Section~\ref{sec:thermo}.
	Multiplying Eq.~\eqref{eq:app_tolman} by $n_i$ then gives $\mathrm{d}P_i/\mathrm{d}r=-(\varepsilon_i+P_i)\,\mathrm{d}\nu/\mathrm{d}r$, the classical content of Eq.~\eqref{eq:TOV_bohm}.

	The quantum term is appended at the fluid level rather than derived alongside the meson forces of Eq.~\eqref{eq:app_force}, and necessarily so: $Q_i^{\mathrm{GR}}$ descends not from the field content of Eq.~\eqref{eq:Lagrangian} but from the gradient expansion of the fermions' own kinetic-energy functional.
	In quantum hydrodynamics that gradient correction contributes the force density $-n_i\,\nabla Q_i$ to the momentum balance \citep{1935ZPhy...96..431W,1957JETP....5...64K,2018PhPl...25c1903M}, with $Q_i$ the Bohm potential; transcribing this force to the static radial geometry and evaluating $Q_i$ with the covariant Laplacian and the effective mass, Eq.~\eqref{eq:bohm}, yields the second term of Eq.~\eqref{eq:TOV_bohm}.
	The metric itself is sourced by the perfect-fluid and meson stress-energy alone, the gradient stress being omitted from Eq.~\eqref{eq:app_einstein}; the omission perturbs the gravitational mass at the order of the integrated surface energy, $\Delta M/M\lesssim 10^{-21}$, which bounds at the same negligible level the violence done to the Bianchi consistency by the appended term.
	The curvature corrections thereby neglected are of the same order as the term discarded in Appendix~\ref{app:closure} and far below any observable considered in this work.

	\section{The closure relation with a vector channel}
	\label{app:closure}

	This appendix derives the first integral~\eqref{eq:closure} from the species-wise force balance~\eqref{eq:TOV_bohm}, established in Appendix~\ref{app:tov}.

	For uniform matter at fixed composition the species-wise relation $\mathrm{d}P_i=n_i\,\mathrm{d}\mu_i$ holds by the construction of Appendix~\ref{app:tov}, with $\mu_i$ given by Eq.~\eqref{eq:mu_i}; at the level of the totals the corresponding identity may be verified directly from Eqs.~\eqref{eq:eps_tot}--\eqref{eq:P_tot} upon using the field equations~\eqref{eq:sigma_alg}--\eqref{eq:omega_alg}, the meson-field variations cancelling between the kinetic and potential terms precisely because the fields are extremal at fixed sources.
	In particular, the often-vexing question of how to apportion the field energies between the species never arises in the chemical-potential formulation: the vector contribution enters $\mu_i$ as the whole term $g_{\omega i}\,\omega_0$, and the scalar contribution enters implicitly through $m_i^*$.

	Dividing Eq.~\eqref{eq:TOV_bohm} by $n_i$ and using $\mathrm{d}P_i=n_i\,\mathrm{d}\mu_i$ together with $\varepsilon_i+P_i=\mu_i n_i$ (which holds by the construction of Appendix~\ref{app:tov}) gives
	\begin{equation}
		\frac{\mathrm{d}\mu_i}{\mathrm{d}r} + \mu_i\,\frac{\mathrm{d}\nu}{\mathrm{d}r} + \frac{\mathrm{d}Q_i^{\mathrm{GR}}}{\mathrm{d}r} = 0\,.
		\label{eq:app_balance}
	\end{equation}
	Equation~\eqref{eq:app_balance} falls short of a total derivative by the single term $Q_i\,\mathrm{d}\nu/\mathrm{d}r$; since $Q_i$ is appreciable only within the surface layer, where it remains far below $\mu_i$, this term is smaller than the retained $\mathrm{d}Q_i^{\mathrm{GR}}/\mathrm{d}r$ by a factor of order $\kappa\,t/R_T\sim\kappa\,m_1/\Mpl$, with $t$ the surface thickness of Section~\ref{sec:bohm_surface}, and may be appended at a cost far below any observable considered in this work; it is the resulting exactified closure, rather than the differential form, that the numerical implementation imposes. With the term included the equation becomes a total derivative,
	\begin{equation}
		\frac{\mathrm{d}}{\mathrm{d}r}\Bigl[\bigl(\mu_i + Q_i^{\mathrm{GR}}\bigr)\,e^{\nu}\Bigr]=0\,,
	\end{equation}
	whence Eq.~\eqref{eq:closure} with $\mu_i^\infty$ the redshifted chemical potential measured at infinity.
	At the surface of a self-bound star, where $e^{\nu(R_T)}=\sqrt{1-2GM/R_T}$ and $Q_i\to 0$ outside the kink, the closure fixes $\mu_i^\infty=\mu_{i,s}\sqrt{1-2GM/R_T}$ with $\mu_{i,s}$ the bulk coexistence chemical potential, which is the boundary condition used throughout.

	\section{The surface kink and the tidal boundary condition}
	\label{app:kink}

	\subsection{First integral and surface tension}

	Within the surface layer the curvature of spacetime and the spherical geometry are both negligible, the layer being thinner than the stellar radius by a factor of order $m_1/\Mpl$; the closure~\eqref{eq:closure} at fixed $\mu_s$ then reads $\mu(\psi^2)+Q=\mu_s$ with $\mu(n)$ the chemical potential of uniform bulk matter, $\psi=\sqrt{n}$, and $Q=-(\lamB/2m_s^*)\,\psi''/\psi$ in the planar coordinate $z$.
	Multiplying by $(\psi^2)'$ and integrating once, using $\mathrm{d}\varepsilon/\mathrm{d}n=\mu(n)$, yields the first integral~\eqref{eq:kink}; the constant of integration vanishes because $w$ and $\psi'$ vanish together in the bulk, and the coexistence condition $w(\nsat)=-P(\nsat)=0$ guarantees that the vacuum boundary condition is met at the other end, the same degeneracy of minima that produces a kink in any double-well problem.
	The profile follows by quadrature,
	\begin{equation}
		z(\psi) = -\int^{\psi}\sqrt{\frac{\lamB}{2m_s^*\,w(\tilde\psi^2)}}\;\mathrm{d}\tilde\psi\,,
		\label{eq:app_profile}
	\end{equation}
	and the surface tension, defined as the excess energy per unit area, is twice the integrated gradient energy,
	\begin{equation}
		\sigma_s = 2\int_{-\infty}^{\infty}\frac{\lamB}{2m_s^*}\left(\frac{\mathrm{d}\psi}{\mathrm{d}z}\right)^{\!2}\mathrm{d}z
		= 2\int_0^{\sqrt{\nsat}}\sqrt{\frac{\lamB\,w(\psi^2)}{2m_s^*}}\;\mathrm{d}\psi\,.
		\label{eq:app_sigma}
	\end{equation}
	Treating $m_s^*$ as constant across the layer at its surface value introduces an error of order ten per cent (the effective mass runs from $0.54\,m_1$ at the bulk edge towards $m_1$ in the vacuum tail, and enters the thickness and tension only under a square root), comparable to the neglected meson-gradient contributions, and entirely adequate for the order-of-magnitude duties the surface tension performs in Section~\ref{sec:surface}.

	\subsection{Tidal Love number of a self-bound star}

	The quadrupolar Love number $k_2$ follows from the logarithmic derivative $y_R=y(R_T)$ of the metric perturbation, obtained by integrating the standard Riccati equation through the background star from the regular-centre condition $y(0)=2$ \citep{2008ApJ...677.1216H,2008PhRvD..77b1502F}.
	For self-bound matter the energy density is discontinuous at the surface, and $y_R$ must be corrected by the standard matching term,
	\begin{equation}
		y_R \;\to\; y_R - \frac{4\pi R_T^3\,\varepsilon_{\mathrm{surf}}}{M}\,,
		\label{eq:app_yR}
	\end{equation}
	exactly as for strange quark stars; omitting the correction overestimates $k_2$ substantially at high compactness.
	The dimensionless deformability then follows from $k_2$ and the compactness exactly as in Section~\ref{sec:tidal}.
	As an internal check, the low-compactness limit of the self-bound configurations, which are nearly uniform spheres, yields $k_2=0.73$ at $\kappa\simeq 10^{-3}$, within a few per cent of the incompressible Newtonian value $3/4$ (the residual reflecting the slight non-uniformity of the profile), whilst the low-compactness free-gas sequence yields $k_2\simeq 0.14$, the Newtonian value for a $\gamma=5/3$ polytrope \citep{2008ApJ...677.1216H}; both limits emerge from the same code path that produces Fig.~\ref{fig:tidal}.

\end{document}